\documentclass[twocolumn]{aa}
\usepackage[varg]{txfonts}
\usepackage{amsmath}
\usepackage{amsfonts}
\usepackage{amssymb}
\usepackage{upgreek}
\usepackage{natbib}
\usepackage{xcolor}
\usepackage{hyperref}
\hypersetup{
    colorlinks=true,
    linkcolor=blue,
    citecolor=blue,
    filecolor=magenta,      
    urlcolor=cyan,
}

\begin{document}

\title{Transition to turbulence in nonuniform coronal loops\\ driven by torsional Alfv{\'e}n waves}
\author{Sergio D\'{i}az-Su\'{a}rez \inst{\ref{inst1},\ref{inst2}}\and Roberto Soler \inst{\ref{inst1},\ref{inst2}}}
\institute{Departament de F\'{i}sica, Universitat de les Illes Balears, E-07122, Palma de Mallorca, Spain \label{inst1} \and Institute of Applied Computing \& Community Code (IAC3), Universitat de les Illes Balears, E-07122, Palma de Mallorca, Spain \label{inst2};  \email{s.diaz@uib.es}}

\date{Received 18 December 2020 /
Accepted 29 January 2021}

\abstract{Both observations and numerical simulations suggest that Alfv\'enic waves may carry sufficient energy to sustain the hot temperatures of the solar atmospheric plasma. However, the thermalization of  wave energy is  inefficient unless very short spatial scales are considered. Phase mixing is a mechanism that can take energy down to dissipation lengths, but it operates over too long a timescale.  Here, we study how turbulence, driven by the nonlinear evolution of phase-mixed torsional Alfv\'en waves in coronal loops, is able to take wave energy down to the dissipative scales much faster than  the theory of linear phase mixing predicts. We consider a simple model of a transversely nonuniform cylindrical flux tube with a constant axial magnetic field. The flux tube is  perturbed by the fundamental  mode of standing torsional Alfv\'en waves. We solved the three-dimensional (3D) ideal magnetohydrodynamics equations numerically to study the temporal evolution. Initially, torsional Alfv\'en waves undergo the process of  phase mixing because of the transverse variation of density. After only few periods of torsional waves, azimuthal shear flows generated by phase mixing eventually trigger the Kelvin-Helmholtz instability (KHi), and the flux tube is subsequently driven to a turbulent state. Turbulence is very anisotropic and develops transversely only to the background magnetic field. After the onset of turbulence, the  effective  Reynolds  number  decreases  in  the  flux tube  much faster than in the initial linear stage governed by phase mixing alone. We conclude that the nonlinear evolution of torsional Alfv\'en waves, and the associated KHi, is a viable mechanism for the onset of turbulence in coronal loops. Turbulence can significantly speed up the generation of small scales. Enhanced deposition rates of wave energy into the coronal plasma are therefore expected.}

\titlerunning{Torsional Alfv{\'e}n waves in nonuniform solar flux tubes}
\authorrunning{S. D\'{i}az-Su\'{a}rez \& R. Soler}
\maketitle

\section{Introduction}

The existence of Alfv\'{e}n waves in magnetized plasmas was first postulated by \citet{Alf42}. In a uniform plasma of infinite extent, Alfv\'{e}n waves are a type of magnetohydrodynamic (MHD) wave of a pure magnetic nature. They are incompressible, their restoring force is magnetic tension, and they are transverse, that is, they are polarized  perpendicularly to the direction of the magnetic field. The energy carried by Alfv\'{e}n waves strictly propagates along the magnetic field direction \citep[see e.g.,][for further details]{Priest12,stix12,Jess15}. In nonuniform plasmas, MHD waves have mixed properties in general and Alfv\'{e}n waves are usually coupled with magnetosonic waves. An example is the so-called kink mode in magnetic flux tubes, which is a transverse MHD wave with a highly Alfv\'{e}nic character \citep{Goossens12}. Torsional Alfv\'{e}n waves, which are pure Alfv\'{e}n waves in straight flux tubes, even in the presence of inhomogenity, are the exception to this rule.

High-resolution and high-cadence observations have shown the ubiquity of transverse MHD waves through the solar atmosphere \citep[see e.g.,][]{Depon07,Jess09,Depon14,Morton15,Jafar17,Srivastava17}.  Some of these observations have been interpreted as Alfv\'{e}n or Alfv\'{e}nic waves. These waves may play a fundamental role in the transport and dissipation of energy. Consequently, they may contribute to the energy balance of the solar corona \citep[see e.g.,][]{Hollweg78,Crammer05,Cargill11,Mathio13,Jess15,Soler19} and to the acceleration of  solar and stellar winds \citep[see e.g.,][]{Charbonneau95,Cranmer09,Matsumoto12,Shoda2018}.

Coronal loops are closed magnetic structures whose footpoints are anchored at the photosphere, where  plasma motions can excite torsional Alfv\'en waves \citep[see, e.g.,][]{fedun2011,Shelyag2011,Shelyag2012,Wedemeyer2012,Mumford2015,Srivastava2017}. \citet{Zaqarashvili2003} suggested that torsional Alfv\'{e}n waves could be detected  by a periodic variation of spectral-line width. The method was used by \citet{Jess09} to detect torsional Alfv\'en waves in a photospheric bright point. \citet{depontieu2012} detected ubiquitous torsional motions in spicules with the Swedish Solar Telescope (SST) that could be related with torsional Alfv\'en waves. More recently, torsional oscillations  at coronal heights were reported by \citet{Kohutova2020} with the  {\em Interface Region Imaging Spectrograph} (IRIS), while \citet{Aschwanden20} detected oscillations in the magnetic free energy during solar flares that are interpreted as torsional Alfv\'{e}n waves.

 Torsional Alfv\'{e}n waves excited at the loop feet can resonate with  standing modes of the loop and drive global torsional oscillations \citep[see, e.g.,][]{hollweg1984}. \citet{Soler20} have investigated the excitation of standing torsional Alfv\'en waves driven by  waves propagating from the photosphere. Due to the presence of cavity resonances, they found that large energy fluxes can be transmitted to the loop by overcoming the filtering effect of the chromosphere. The transmission of energy mostly occurs at a frequency corresponding to the fundamental standing torsional mode of the loop.  \citet{Soler20} concluded that the transmitted energy is not efficiently dissipated in the corona. However,   \citet{Soler20} only considered the linear regime.

Torsional Alfv\'{e}n waves in transversely nonuniform magnetic flux tubes undergo the process of phase mixing, which is linear in nature \citep[see, e.g.,][]{HeyvaertPriest83,Nocera84,Moortel00,Smith07,Prokopyszyn2019}. As first shown by \citet{HeyvaertPriest83} and \citet{Browning}, shear flows generated during the phase mixing evolution can nonlinearly trigger the Kelvin-Helmholtz instability (KHi). The KHi has been observed in coronal mass ejections and quiescent prominences \citep[see,  e.g.,][]{Berger10,Ryutova10, Foullon11,Ofman11,hillier2018}. Nevertheless, there are no direct observations of the KHi on coronal loops. The nonlinear evolution of the KHi can induce the formation of eddies and the transition to turbulence, as numerical simulations suggest \citep[see, e.g.,][]{Terradas08,Antolin15,Magyar16,Howson17,Terradas18,Karampelas19,Antolin19}. These previous works have studied kink oscillations, while the case of torsional oscillations has only been explored by \citet{Guo19}. Turbulence  might significantly enhance the efficiency of wave heating of the coronal plasma by rapidly cascading energy from large scales to the dissipative scales \citep[][]{Hillier2020}. Indeed,  there is evidence that coronal loops may be in a turbulent state \citep{DeMoortel2014,Liu2014,Hahn2014}. However, there has not yet been any direct observational connection between torsional Alfv\'{e}n waves, the associated KHi, and the alleged turbulence. Here, we numerically investigate this mechanism for the driving of turbulence in coronal loops.

The paper is organized as follows. Section 2 describes the background model and the numerical set-up. In Sect. 3, we include a quasi-linear theoretical analysis where we apply perturbation theory. Some thoughts about excitation of the KHi are also given. Results from the full numerical simulations are shown in Sect. 4. Finally, the conclusions of this work are discussed in Sect. 5.

\section{Setup}

\subsection{Model}

\begin{figure}[htpb]
\centering
\includegraphics[width=0.75\columnwidth]{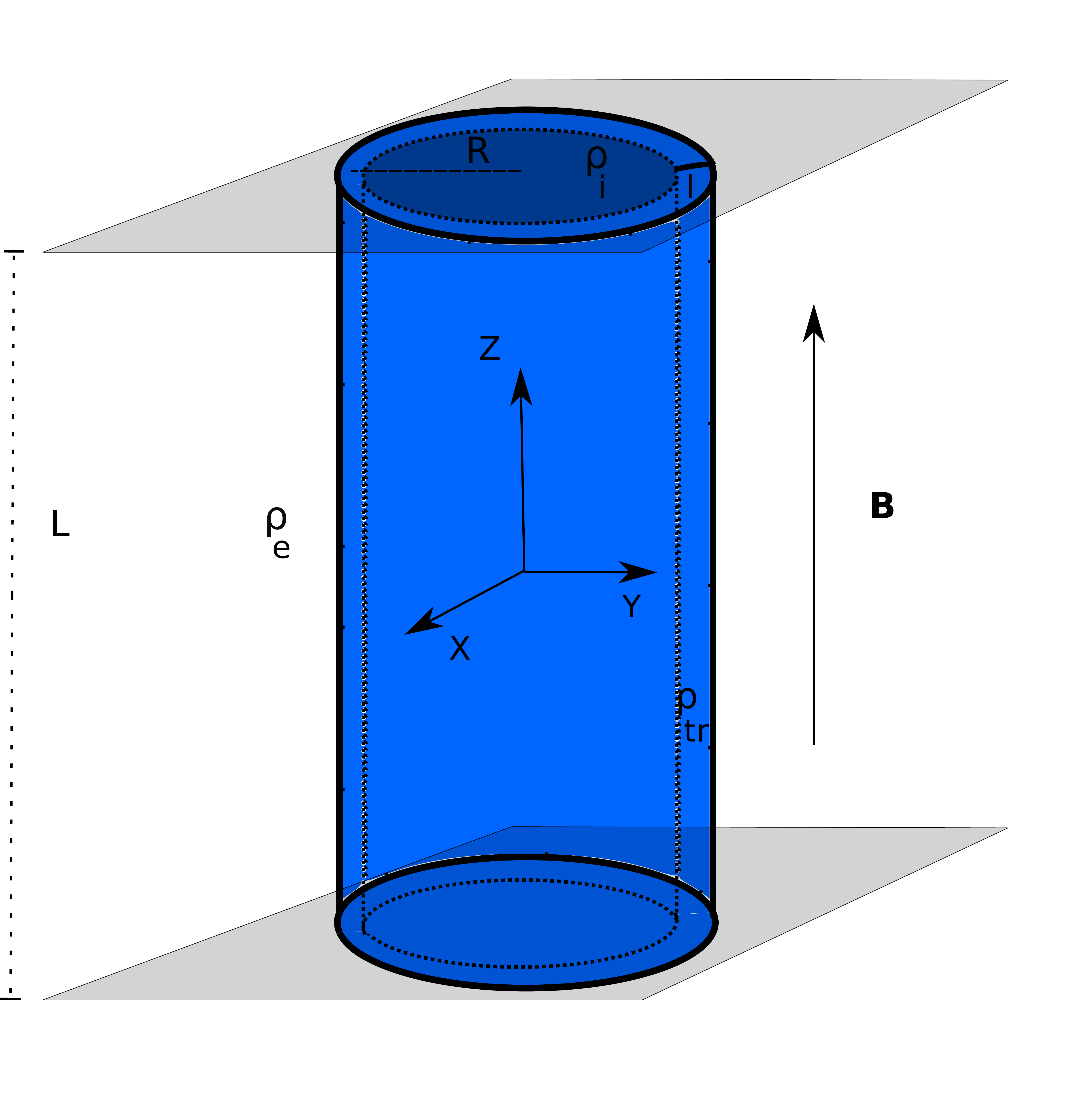}
\caption{Schema of the coronal flux tube model. The two gray planes located at the ends of the tube represent the solar photosphere where the magnetic field is anchored.}
\label{scheme1}
\end{figure}

{To represent a coronal loop, we considered the so-called standard  flux tube  model frequently used in the wave literature \citep[see, e.g.,][]{Edwin83}.} The model is made of a cylindrical tube of length $ L $ and radius $ R $, embedded in a uniform coronal environment. We used a reference frame so that the $ z$-direction would point along the axis of the flux tube. The magnetic field is straight, longitudinal to the flux tube, and uniform everywhere, namely $ \mathbf{B}=B_{0}\hat{z}$, {with $B_0$ the magnetic field strength}. The loop footpoints are line-tied at two rigid walls representing the {much denser} solar photosphere. So, the present model neglects the curvature of the coronal loop {and the thin chromospheric layer at the loop feet}. A schema of the model can be seen in Fig. \ref{scheme1}.

\begin{figure}[htbp]
\centering
\resizebox{\hsize}{!}{\includegraphics{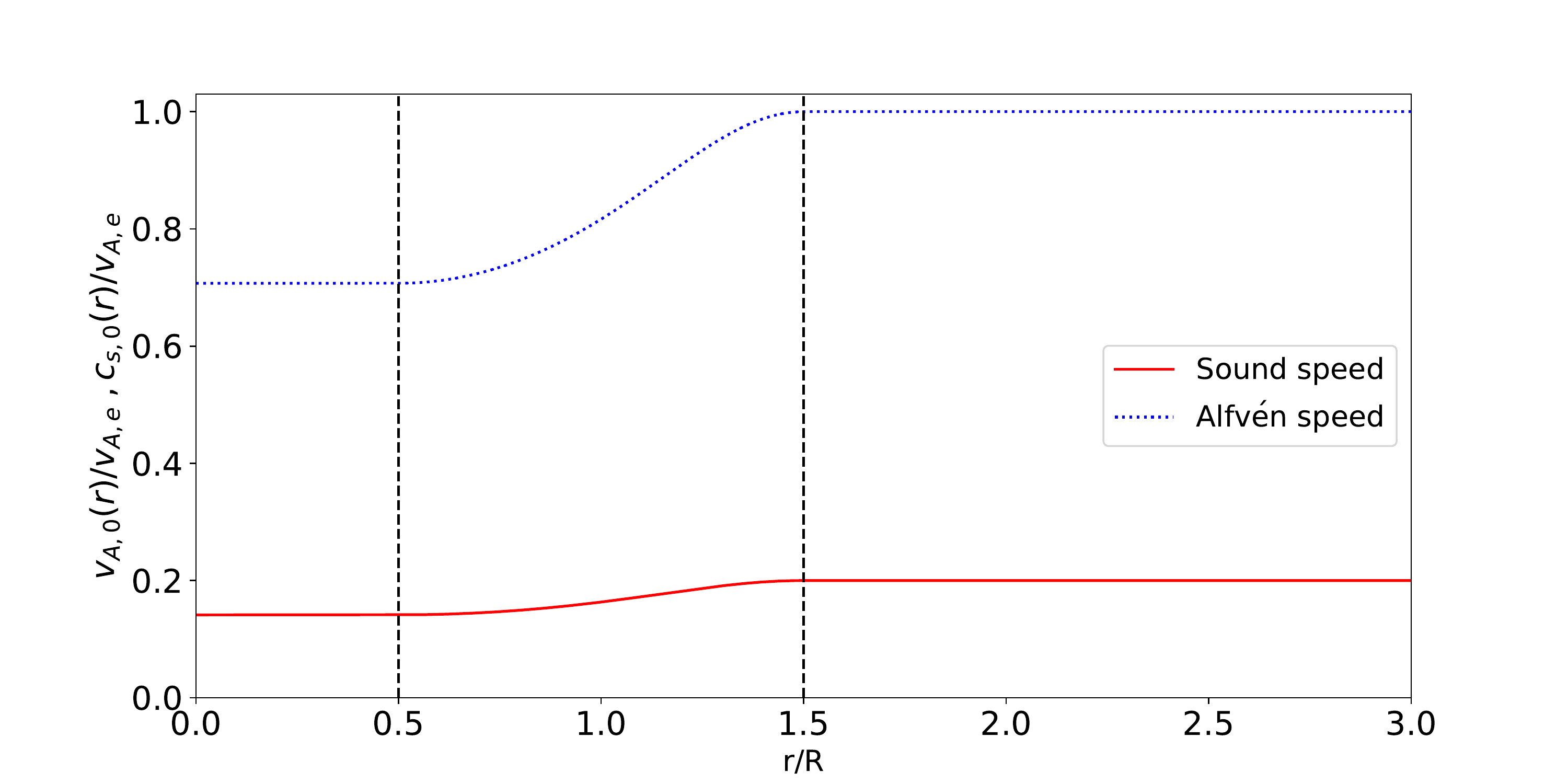}}
\caption{Radial profiles of equilibrium Alfv\'{e}n and sound speeds in a loop with $l/R=1$ and $\rho_{i}/\rho_{e}=2$. The values are normalized with respect to the external Alfv\'{e}n speed. The vertical dashed lines denote the boundaries of the nonuniform layer.} 
\label{velpro}
\end{figure}

The coronal loop equilibrium density is nonuniform in the transverse direction only, namely
\begin{equation}
\rho_0(r)=
\left\{
        \begin{array}{lll}
                \rho_{\mathrm{i}},  & \mbox{if } & r \leq R-\frac{l}{2}, \\
                \rho_{\mathrm{tr}}(r),  & \mbox{if } &  R-\frac{l}{2}< \; r  < R+\frac{l}{2}, \\
                \rho_{\mathrm{e}},  & \mbox{if } & r \geq R+\frac{l}{2},
        \end{array}
\right.
\label{eqdensity}
\end{equation}
where $ r $ is the radial coordinate, $\rho_{\mathrm{i}} $ is the density in the uniform inner core, $\rho_{\mathrm{e}} $ is the uniform external density, and $ \rho_{\mathrm{tr}}(r) $ is the density in a nonuniform transitional layer of thickness $ l $ that continuously connects both uniform regions as
\begin{equation}
\rho_{\mathrm{tr}}(r)=\frac{\rho_{\mathrm{i}}}{2}\left\{\left[1+\frac{\rho_\mathrm{e}}{\rho_{\mathrm{i}}}\right]-\left[1-\frac{\rho_\mathrm{e}}{\rho_{\mathrm{i}}}\right]\sin\left[\frac{\pi}{l}\left(r-R\right)\right]\right\}.
\label{rhotr}
\end{equation}
The allowed values of $ l $ range  from $ l=0 $ (abrupt transition) to $ l=2R$ (fully nonuniform loop). We set the equilibrium gas pressure to be a constant everywhere, namely $p_0$, so that the plasma $ \beta = \frac{p_0}{B_0^2/\mu}    \approx 0.024$, where $\mu$ is the magnetic permeability. {The radial profiles of the equilibrium Alfv\'{e}n speed, $ v_{A,0}(r) = B_{0}/\sqrt{\mu\rho_0(r)}$, and the equilibrium sound speed, $ c_{s,0}(r) = \sqrt{\gamma p_{0}/\rho_0 (r)}$, where $\gamma$ is the adiabatic constant,  are displayed in Fig. \ref{velpro}  for a particular set of parameters.}

\subsection{Numerical code}
\label{subsect:numcode}

We used the PLUTO code \citep{Mignone07} to solve the three-dimensional (3D) ideal MHD equations with a finite-volume, shock capturing spatial discretization on a structured mesh. The equations are as follows:
\begin{align}
\frac{\partial\rho}{\partial\mathrm{t}}&=-\nabla \cdot (\rho \mathbf{v}), \label{con1}\\
\mathrm{\rho \frac{D \mathbf{v}}{Dt}}&=-\vec{\nabla}p+\frac{1}{\mu}\left(\nabla \times \mathbf{B}\right)\times \mathbf{B},\label{mom1}\\
\frac{\partial \mathbf{B}}{\partial \mathrm{t}}&=\nabla \times (\mathbf{v}\times \mathbf{B}),\label{ind1} \\
\frac{\mathrm{D}p}{\mathrm{D}t}&=\frac{\upgamma p}{\rho}\frac{\mathrm{D}\rho}{\mathrm{D}t}.\label{ene1}
\end{align}
In these equations, $ \mathrm{\frac{D}{Dt}}\equiv \frac{\partial}{\partial t}+ \mathbf{v}\cdot \nabla $ denotes the total derivative, $ \rho $ is the mass density, $ \mathbf{v} $ is the velocity, $ p $  is the gas pressure, and $ \mathbf{B} $ is the magnetic field. Gravity and nonideal terms are neglected.

The code solves Eqs. (\ref{con1})-(\ref{ene1}) in Cartesian coordinates. We used a Roe-Riemann solver \citep{Roe81} to compute the numerical fluxes, a second-order parabolic scheme for spatial reconstruction, and a second-order Runge-Kutta algorithm for temporal evolution. The extended GLM method \citep{Dedner02} was employed to keep the solenoidal constraint on the magnetic field. Moreover, an adaptive mesh refinement (AMR) strategy was used \citep{Mignone12}. The use of AMR allows us to decrease the  computational cost by using a high resolution in the region of interest while keeping a lower resolution far from the relevant dynamics. The computational box has a base numerical resolution of 100x100x100 cells, which are distributed uniformly from  $-3R$ to $3R$ in the  $x$-direction and the $y$-direction, and from $-L/2$ to $L/2$ in the $z$-direction. We included four levels of refinement in which the cells double the resolution from one level to the next so that the maximum effective resolution is 1600x1600x1600. {For typical values of the loop radius, the effective transverse resolution is  $\sim 10$~km}. The PLUTO criterion for refinement is based on the second derivative error norm \citep{Lohner87} of the perturbation of the total energy, that is, the total energy of the system less the internal and magnetic energy of the background.

Regarding the boundary conditions, we consider outflow conditions, that is, zero gradient, for the pressure, density, the $x$- and $y$-components of the magnetic field, while zero velocities are imposed at all boundaries. Concerning the conditions for the $z$-component of the magnetic field, $ B_{z} $, we also considered outflow conditions at the lateral boundaries, but $ B_{z} $ is fixed  to the equilibrium  value, $ B_{0} $, at the bottom and top boundaries, that is, at $ z=\pm L/2$, in order to implement the line-tying of the field lines at the photosphere. With these boundary conditions, we checked the nonexistence of significant reflections at the lateral boundaries of the domain during the simulations. Furthermore, a background splitting technique \citep{Powell94} was used, which has the advantage that {only the magnetic field perturbation over the background magnetic field is evolved by the code}.

{A recent investigation by \citet{Soler20} shows that coronal loop torsional oscillations excited by a photospheric broadband driver are dominated by the longitudinal fundamental mode of the loop. In view of this result, we} imposed an initial condition for the velocity that aims to excite  the longitudinally fundamental mode of standing torsional Alfv\'{e}n waves. To do so, we imposed the following purely azimuthal velocity field at $t=0$ :
\begin{equation}
\mathbf{v}\left(t=0\right)= v_{0}A(r)\cos\left(\frac{\pi}{L}\;z\right)\; \hat{\varphi},
\label{vazi}
\end{equation}
where $ v_{0} $ is the maximum velocity amplitude and $ A(r)$ contains the radial dependence. We shall use the particular form $ A(r)=A_{0}r\exp{\left[-\left(r^{2}/\upsigma^{2}\right)\right]} $ with $ \upsigma=0.9R $ and $ A_{0}=\exp(0.5)\sqrt 2/\upsigma $ is a constant that we obtain from imposing that the maximum value of the velocity is equal to the prescribed value, $ v_{0} $. {We note that most of the results discussed in this paper would remain practically unchanged if a different dependence for $ A(r)$ was used.}

\section{Quasi-linear theoretical analysis}

{Before analyzing the results of} the full numerical computations, the purpose here is to solve Eqs. (\ref{con1})-(\ref{ene1}) in an analytic approximated manner when the standing torsional Alfv\'{e}n waves behave quasi-linearly. These approximate solutions can help us understand the full nonlinear evolution. To do so, we applied a regular perturbation theory \citep[see, e.g.,][]{Pep20}. {In the case of propagating torsional Alfv\'{e}n waves, a weak nonlinear analysis using the second order thin flux tube approximation can be found in \citet{Shestov17}.}

We define the parameter $ \varepsilon \equiv v_{0}/v_{A,e} $ where $ v_{A,e}$ is the external Alfv\'{e}n speed. Then, assuming $ \varepsilon<<1 $, we write
\begin{align}
\rho&=\rho_{0}+\varepsilon^{2}\rho', \nonumber  & \mathbf{v}&=\varepsilon v_{\varphi}' \hat{\varphi}+\varepsilon^{2} v_{z}' \hat{z},\nonumber  \\
 p&=p_{0}+\varepsilon^{2}p',\nonumber  & \mathbf{B}&=\varepsilon B_{\varphi}' \hat{\varphi}+B_{0} \hat{z},\nonumber 
\nonumber
\end{align}
where the subscript 0 denotes a background quantity, and the prime $'$ denotes a small perturbation. We substituted these expressions into Eqs. (\ref{con1})-(\ref{ene1}) and separated the terms according to their order in $ \varepsilon $. The first-order equations in $ \varepsilon$ govern $ v_{\varphi}' $, and $ B_{\varphi}'$, so they describe linear torsional Alfv\'{e}n waves. The second-order equations in $ \varepsilon^{2} $ govern $ p' $, $ \rho ' $, and $ v_{z}'$ and  describe  nonlinear effects as the ponderomotive force, and the  coupling of Alfv\'{e}n waves with slow magnetoacoustic  waves.

\subsection{First-order equations: Transverse dynamics}

By combining the first-order equations for $ v'_{\varphi} $ and $ B_{\varphi}'$, we can arrive at an equation involving $ v_{\varphi}' $ alone, namely
\begin{equation}
\frac{\partial^{2}v_{\varphi}'}{\partial t^{2}}=v^{2}_{A,0}(r)\frac{\partial^{2}v_{\varphi}'}{\partial z^{2}},
\label{torsional}
\end{equation}
where  $ v_{A,0}(r) $ is the equilibrium Alfv\'{e}n velocity, which we recall is a function of the radial direction. Equation~(\ref{torsional}) is the 1D wave equation with phase velocity equal to the Alfv\'{e}n speed. We note that in our formalism the actual amplitude of the azimuthal velocity perturbation is $ \varepsilon v_{\varphi}'$. Equation~(\ref{torsional}) can be solved for the longitudinally fundamental mode by taking into account the prescribed boundary and initial conditions. The solution is
\begin{equation}
\varepsilon v_{\varphi}'=v_{0}A(r)\cos\left(\frac{\pi}{L}\; z\right)\cos[\omega_{A}(r) t],
\label{vphi}
\end{equation}
where $ \omega_{A} (r)=\frac{\pi}{L}v_{A,0}(r) $ is the radially-dependent Alfv\'{e}n frequency. In turn, the azimuthal magnetic field perturbation is
\begin{equation}
\varepsilon B_{\varphi}'=-B_{0}\frac{v_{0}}{\omega_{A}(r)}\frac{\pi}{L} A(r)\sin\left(\frac{\pi}{L}\; z\right)\sin[\omega_{A}(r) t].
\label{bphi}
\end{equation}

\begin{figure}[htpb]
\centering
\resizebox{\hsize}{!}{\includegraphics{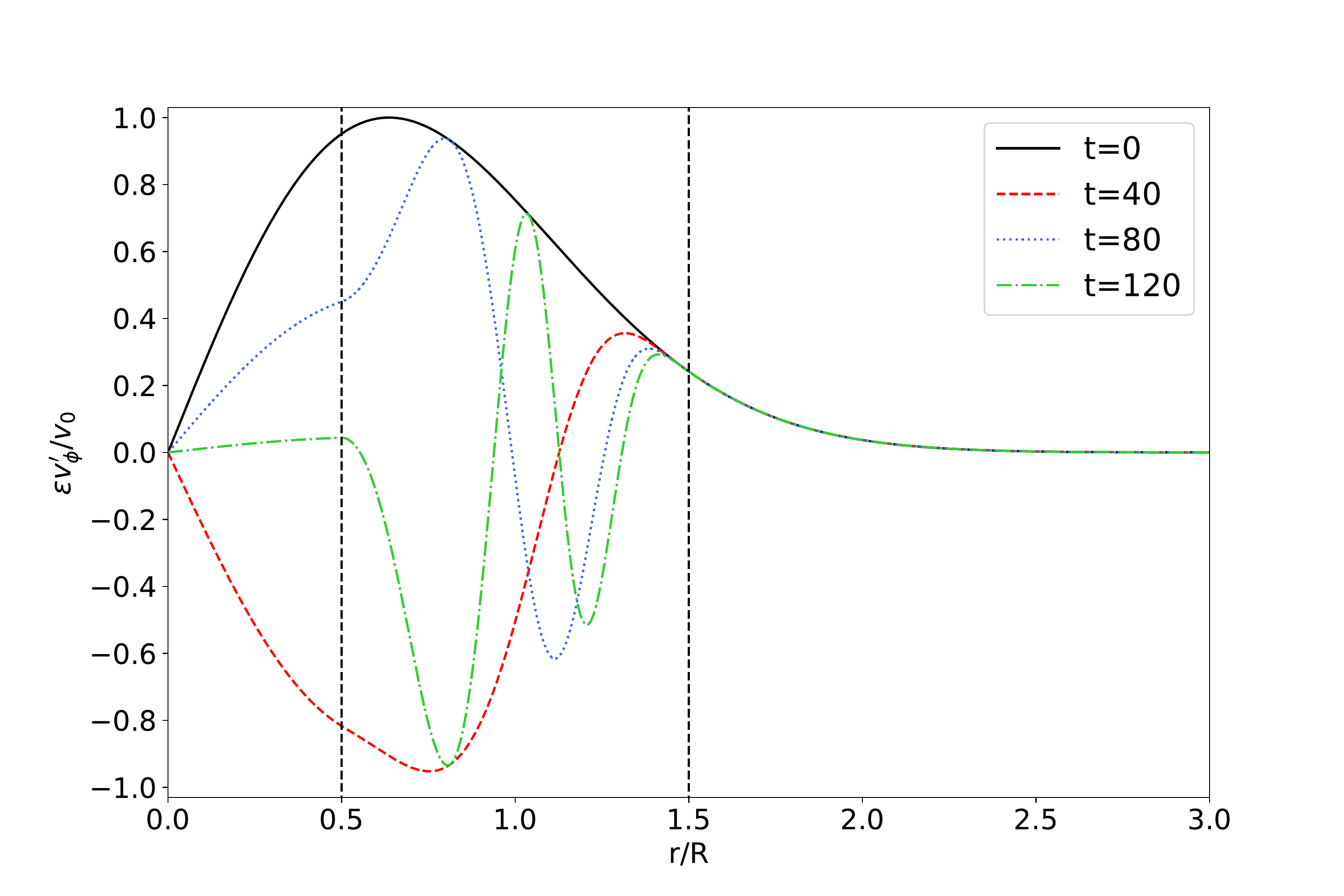}}
\caption{Azimuthal component of the velocity perturbation normalized with respect to the initial amplitude as a function of the radial position at various times. We used $l/R=1 $, $ \rho_{i}/\rho_{e}=2 $, $L/R=10, $ and $ z=0$. Time is normalized as $ t v_{A,e}/R$. The vertical dashed lines denote the boundaries of the nonuniform layer.}
\label{radiallo}
\end{figure}

Equations~(\ref{vphi}) and (\ref{bphi}) evidence that standing torsional Alfv\'{e}n waves excited by the initial perturbation oscillate with the local Alfv\'{e}n frequency. In the transitional layer, the frequency is nonuniform across the coronal loop. As a result, waves living on adjacent radial positions, excited in phase by our initial condition, will become out of phase as time passes. This is the well-known process of phase mixing \citep[see e.g.,][]{HeyvaertPriest83,Nocera84,Moortel00,Smith07}. Phase mixing continuously decreases the length scale of the disturbances across the loop. In Fig. \ref{radiallo}, we illustrate how the azimuthal velocity perturbation is affected by phase mixing in the inhomogeneous region of the flux tube with time. The transverse length scales of the perturbation in the homogeneous inner core and the homogeneous external plasma remain the same at all times. However, the length scale decreases with time in the transitional layer.

Following \citet{Mann95}, the effective wave number across the loop can be estimated by
\begin{equation}
k_{r} (r) \approx \frac{\partial \omega_{A}(r)}{\partial r}t.
\label{effwnu}
\end{equation}
With time, the effective wave number increases. {In ideal MHDs, this process works indefinitely. However, in dissipative MHDs,  dissipation becomes important for sufficiently high wave numbers at sufficiently long times. Thus, phase mixing transports energy from large to small scales until those small scales can be efficiently dissipated \citep[see, e.g.,][]{Ebrahimi20}. Here, we cannot study the dissipation phase because we restricted ourselves to ideal MHDs.}

\subsection{Second-order equations: Longitudinal dynamics}

In a similar way to what was described in the previous subsection, we combined the second-order equations for $ v_{z} '$, $ \rho'$, and $ p'$ to obtain an equation for the $z$-component of the velocity perturbation, namely
\begin{equation}
\frac{\partial^{2}v_{z}'}{\partial t^{2}}-c^{2}_{s,0}(r)\frac{\partial^{2}v_{z}'}{\partial z^{2}}=-\frac{1}{2\mu\rho(r)}\frac{\partial^{2}B'^{2}_{\varphi}}{\partial t \partial z},
\label{slow}
\end{equation}
where $ c_{s,0}(r) $ is the equilibrium sound speed. Equation (\ref{slow}) is the 1D wave equation with phase velocity equal to the sound speed. We note again that the actual amplitude of the longitudinal velocity perturbation is $ \varepsilon^{2}v_{z}'$. We see the presence of a force term on the right-hand side of Eq. (\ref{slow}), which depends on the perturbation of the magnetic pressure associated with the Alfv\'{e}n waves. {The general solution to Eq.~(\ref{slow}) is the sum of the particular and the homogeneous solutions. The homogeneous solution is physically related to the slow magnetoacoustic waves that can propagate freely along the loop at the sound speed. We are not interested in this solution. Conversely, the particular solution is related to the nonlinear longitudinal dynamics associated with the Alfv\'{e}n waves. In the  case that  $ B_{\varphi}' $ is given by Eq.~(\ref{bphi}), the particular solution to Eq.~(\ref{slow}) can be found by direct integration \citep[see, e.g.,][]{David18}, namely
\begin{eqnarray}
\varepsilon^{2}v_{z}'&=&-v^{2}_{0}\frac{A^{2}(r)}{8c_{s,0}(r)}\left[\frac{v^{2}_{\rm A,0}(r)}{v^{2}_{\rm A,0}(r)-c^{2}_{s,0}(r)}\right]\sin{\left(\frac{2\pi}{L} z\right)} \nonumber \\
 &&\times \Bigg\{\sin\left[\frac{2\pi}{L} c_{s,0}(r) t\right]-\frac{c_{s,0}(r)}{v_{\rm A,0}(r)}\sin\left[2\omega_{\rm A}(r)t\right]\Bigg\}.
\label{vz}
\end{eqnarray}
}
Considering the low-$ \beta $ plasma regime so that $ c_{s,0} (r) <<v_{A,0} (r)$, Eq. (\ref{vz}) can be simplified as
\begin{equation}
\varepsilon^{2} v_{z}' \approx -v^{2}_{0}\frac{A^{2}(r)}{8c_{s,0}(r)}\sin{\left(\frac{2\pi}{L}\; z\right)}\sin\left[\frac{2\pi}{L} c_{s,0}(r) t\right].
\label{vz2}
\end{equation}

\begin{figure}[htbp]
\centering
\resizebox{\hsize}{!}{\includegraphics{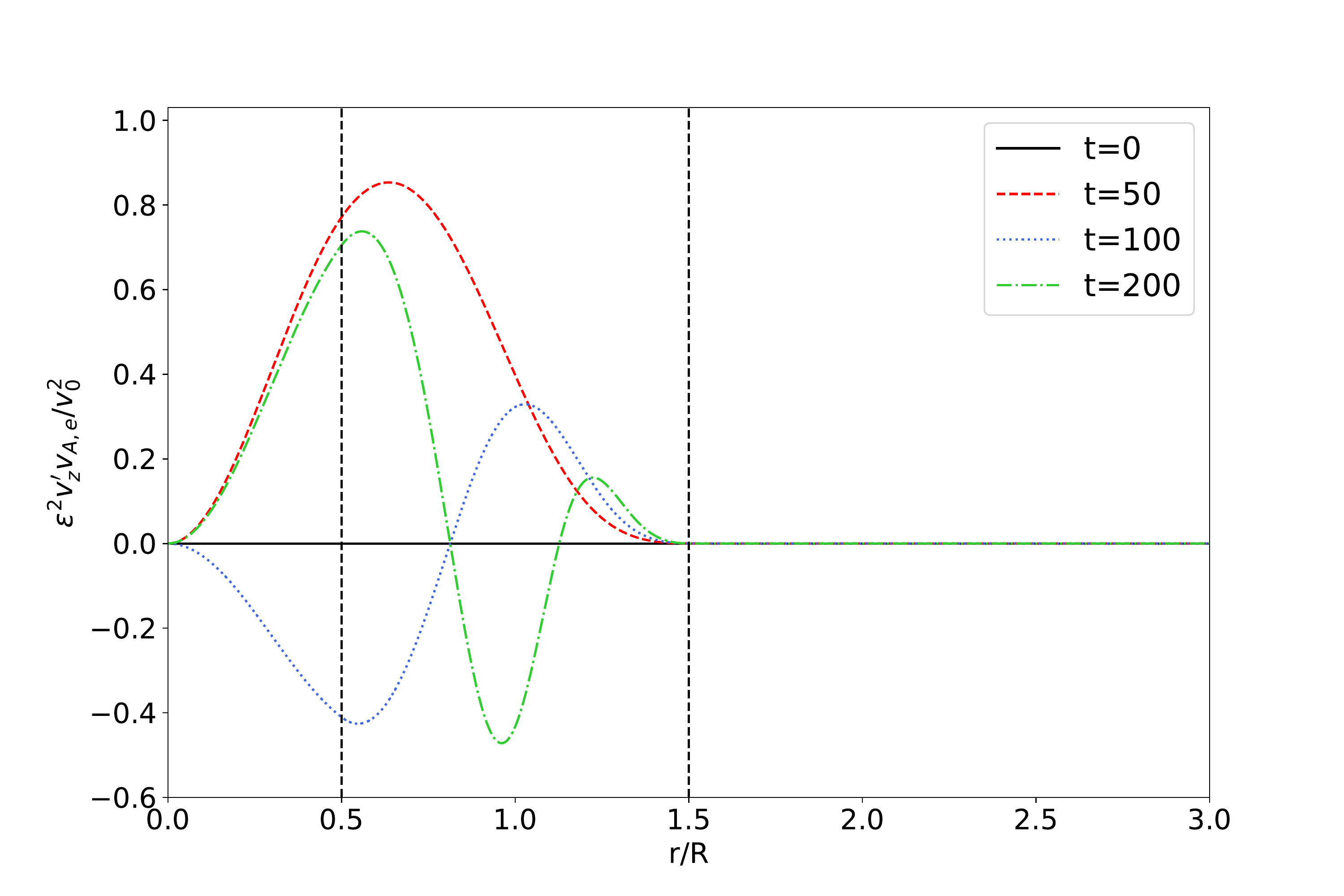}}
\caption{Vertical component of velocity perturbation, normalized as  $ v_{\rm A,e}/v^{2}_{0}$, as a function of the radial position at various times. We have used $l/R=1 $, $ \rho_{i}/\rho_{e}=2 $ $L/R=10 $ and $ z=L/4$. Time is normalized as $ t v_{\rm A,e}/R$. The vertical dashed lines denote the boundaries of the nonuniform layer.}
\label{vzgraph}
\end{figure}

Equation (\ref{vz2}) evidences the existence of longitudinal motions along the flux tube. The amplitude of the longitudinal velocity perturbation depends quadratically on the initial amplitude of the azimuthal velocity, which points out the nonlinear nature of these longitudinal motions. There is a periodic converging (diverging) flow toward (away) from the center of the tube. The result of these flows is the periodic compression and expansion of material around the center of the tube. This is caused by the well-known ponderomotive force \citep[see e.g.,][]{Hollweg71,Rankin94,Tikhonchuk95,Terradas04}. The periodicity of these longitudinal motions is $ L/c_{s,0}(r) $, so it depends on the radial position. Thus, the vertical flows associated with the ponderomotive force also become out of phase in adjacent radial positions of the nonuniform layer as time increases. Figure \ref{vzgraph} displays the longitudinal velocity normalized as  $ \varepsilon^{2} v_{z}' v_{\rm A,e}/v^{2}_{0}$. With this normalization, the dependence on the initial amplitude is dropped, but the radial structure of the longitudinal flow is kept. For sufficiently long times, a longitudinal velocity shear across the tube in the nonuniform layer is evident. 

\subsection{Kelvin-Helmholtz instability: Considerations}

The above quasi-linear analysis indicates that the evolution of the torsional Alfv\'{e}n waves, through phase mixing, results in the occurrence of azimuthal and longitudinal velocity shears in the nonuniform transitional layer. It is well known that a velocity shear in a plasma can drive the KHi \citep[see][]{Chandra}.

The azimuthal shear flows generated by phase mixing can easily develop the KHi, because the velocity shear is perpendicular to the direction of the background magnetic field \citep[see, e.g.,][]{HeyvaertPriest83,Browning,soler2010,Zaqarashvili2015,Barbulescu19}. However, the longitudinal shear flows associated with the ponderomotive force cannot develop the KHi since the velocity shear is along the magnetic field. The effect of magnetic tension prevents the instability of sub-Alfv\'enic flows \citep{Chandra}. Therefore, only the azimuthal shear flows can trigger the KHi in our model.

It is well known by theoretical studies and numerical simulations that the phase-mixing-driven KHi leads to a faster generation of small spatial scales than what the theory of linear phase mixing (Eq.~(\ref{effwnu})) predicts \citep[see, e.g.,][]{Browning,Pagano18,Howson20,Vandamme20}. So, the nonlinear triggering of the KHi can accelerate the energy cascading  to the dissipative scales compared with the effect of linear phase mixing alone.

A relevant question that arises is how to determine the onset time of the KHi. In the case that the velocity shear is not steady but periodic, the mathematical analysis is rather involved \citep[see e.g.,][]{Kelly1965,Roberts1973,Browning,Hillier19,Barbulescu19}. By assuming strong phase mixing, that is, assuming that the exponential growth time of the KHi is much shorter than the period of the oscillating shear flow, \citet{Browning} provided an approximate expression for the critical onset time. The analysis of \citet{Browning} was done in Cartesian geometry, but it can also be applicable to the cylindrical case studied here. In our model parameters, the expression reads
\begin{equation}
t_{\rm crit} =\frac{4 l}{A(R) v_0}, 
\label{tcri}
\end{equation}
where $ A(R) \approx 0.8$ is calculated at a reference point located at the center of the transitional layer, $r=R$. Equation~(\ref{tcri}) indicates that the wider the nonuniform layer, the larger the onset time, as is consistent with the fact that phase mixing develops slower in smooth profiles than in sharp ones \citep{HeyvaertPriest83}. In turn, the onset time is inversely proportional to the velocity amplitude of the initial perturbation, which means that the nonlinear triggering of the KHi would occur faster for large amplitudes than for small amplitudes.

Assuming typical values of the coronal Alfv\'en speed, and the loop length as $v_{\rm A,e} = $~1,000~km~s$^{-1}$, and $L=10^5$~km, the period of the torsional Alfv\'en wave is $P = 2L/v_{\rm A,e} \approx 3.33$~min. In turn, considering a loop radius of $R=$~3,500~km and assuming that the thickness of the nonuniform layer is of the same order, Eq. ~(\ref{tcri}) gives $t_{\rm crit} \approx 2.92$~min for a velocity amplitude of $v_0 = $~100~km~s$^{-1}$ \citep{Kohutova2020}. This simple numerical example indicates that the KHi can be triggered in the loop in a timescale comparable with the period of the torsional oscillations. This implies that the KHi should have a deep impact on the full nonlinear evolution.

\section{Numerical simulations}
\label{Sect:numsimu}
Our aim was to investigate the nonlinear evolution of the phase mixing of torsional Alfv\'en waves and the triggering of the KHi. We focused on studying how these two mechanisms affect the  development of the energy cascade to small scales. Results from the full nonlinear simulations were compared to those from the quasi-linear analytical theory.

Unless otherwise stated, the following reference parameters are used in all simulations: $\varepsilon=0.1 $, $ \rho_{i}/\rho_{e}=2 $ and $L/R=10 $, where $\varepsilon$ is related to the initial velocity amplitude as $v_0 = \varepsilon  v_{A,e}$. We are aware that the considered value of $L/R$ results in a shorter coronal loop than the loop lengths typically  reported in observations. The reason for considering a shorter loop is to speed up the simulation times, since the periods of the torsional oscillations are proportional to $L$. Considering a longer loop would result in longer simulation times, but the dynamics discussed here would be the same for all practical purposes. The effect of the loop length on the triggering of the KHi is explored in Sect.~\ref{parstudy}.

Concerning the thickness of the nonuniform layer, we considered two cases: a thin-layer case with $l/R=0.4, $ and a thick-layer case with $l/R=1.5 $. We considered these two cases because the process of phase mixing, as linear theory predicts, develops at a different pace depending on the inhomogeneity length scale. This is expected to have a strong impact on the nonlinear evolution, including the triggering of the KHi.

In the numerical code, lengths and velocities are normalized with respect to $R$ and $v_{A,e}$, respectively. Density is normalized with respect to the external density. In turn,  time is  normalized with respect to the transverse Alfv\'enic travel time, $R/v_{A,e}$.  In these normalized units, the periods of the internal and external Alfv\'en waves are $20\sqrt{2} \approx 28.3$ and 20 time units, respectively.

The maximum simulation time is determined by the ability of the four-level AMR scheme to correctly describe the small spatial scales that are generated in the system. To check that, we monitored the total energy integrated in the whole computational domain. We stopped the simulation when the integrated total energy started to decrease significantly, meaning that the developed small scales are beyond the maximum effective resolution of the AMR scheme, and numerical dissipation is becoming significant. By stopping the simulation at that point, we ensured that the simulated dynamics was physically meaningful and minimized the influence that undesirable numerical effects might have. The maximum time that can be correctly simulated according to the energy criterion also coincides with the saturation and subsequent unphysical decrease of the vorticity squared and the current density squared integrated over the whole computational box (this is discussed later, in Sect.~\ref{sec:vort}). For the reference parameters given above, the maximum simulation times in the thin-layer and thick-layer cases are $t=86$ and $t=150$, respectively.

\begin{figure*}[!htbp]
\centering
\includegraphics[width=1.85\columnwidth]{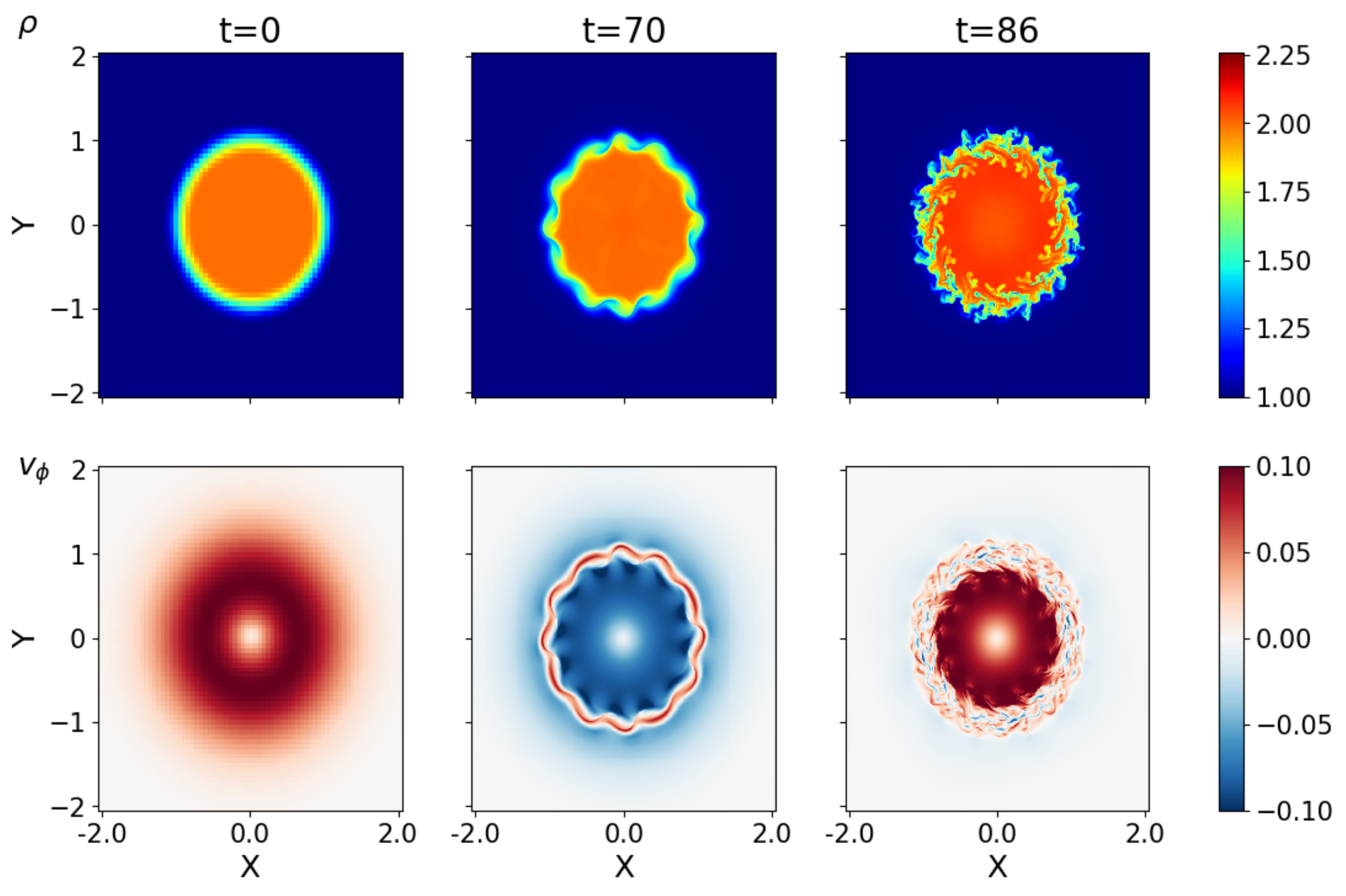}
\caption{\textit{Top}: Cross-sectional cut of density at the tube center, $z=0$, for the thin-layer case at three different simulation times indicated on top of each panel. \textit{Bottom}: Same cut, but for the azimuthal component of velocity. The full temporal evolution is available as an online movie.}
\label{denscutthin}
\end{figure*}

\begin{figure*}[!htbp]
\centering
\includegraphics[width=1.85\columnwidth]{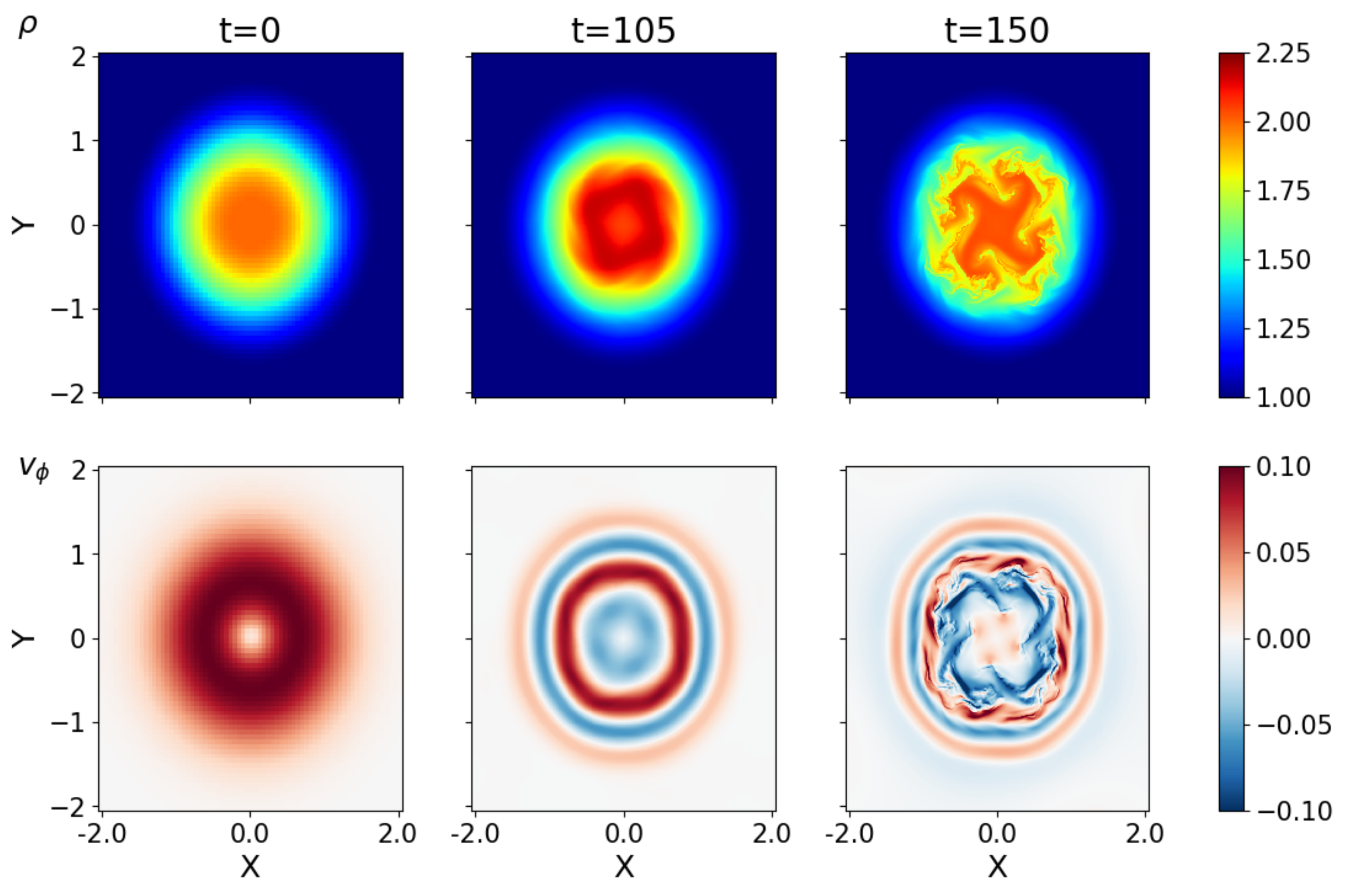} 
\caption{Same as Fig.~\ref{denscutthin}, but for the thick-layer case. We note that the three selected simulation times are different from those of Fig.~\ref{denscutthin}. The full temporal evolution is available as an online movie.}
\label{denscutthick}
\end{figure*}

\subsection{Nonlinear evolution of phase mixing}
\label{sect:rhovphi}

Figures~\ref{denscutthin} and \ref{denscutthick} show the temporal evolution of density and the azimuthal component of  velocity in a cross-sectional cut at the tube center in the thin-layer and thick-layer cases, respectively. Since the relevant physics is inside and near the flux tube, these figures and the following ones only display cross-sectional cuts in a subdomain where $x,y \in \left[ -2R, 2R  \right]$, while we recall that the complete numerical domain is larger with  $x,y \in \left[ -3R, 3R  \right]$. At the beginning of the simulations, the wave evolution agrees with the predictions of quasi-linear theory. As time passes, the alternation between red (positive) and blue (negative) values of the azimuthal component of  velocity within the nonuniform layer is a clear evidence that the process of phase mixing is at work \citep[see][]{HeyvaertPriest83}. In the transition region, where the density is nonuniform, adjacent radial positions become out of phase as the simulation evolves, generating  azimuthal shear flows and smaller spatial scales across the tube. The development of phase mixing is slower in the thick-layer case than in the thin-layer case because of the smoother Alfv\'{e}n  speed gradient.

\begin{figure*}[!htbp]
\centering
\includegraphics[width=1.85\columnwidth]{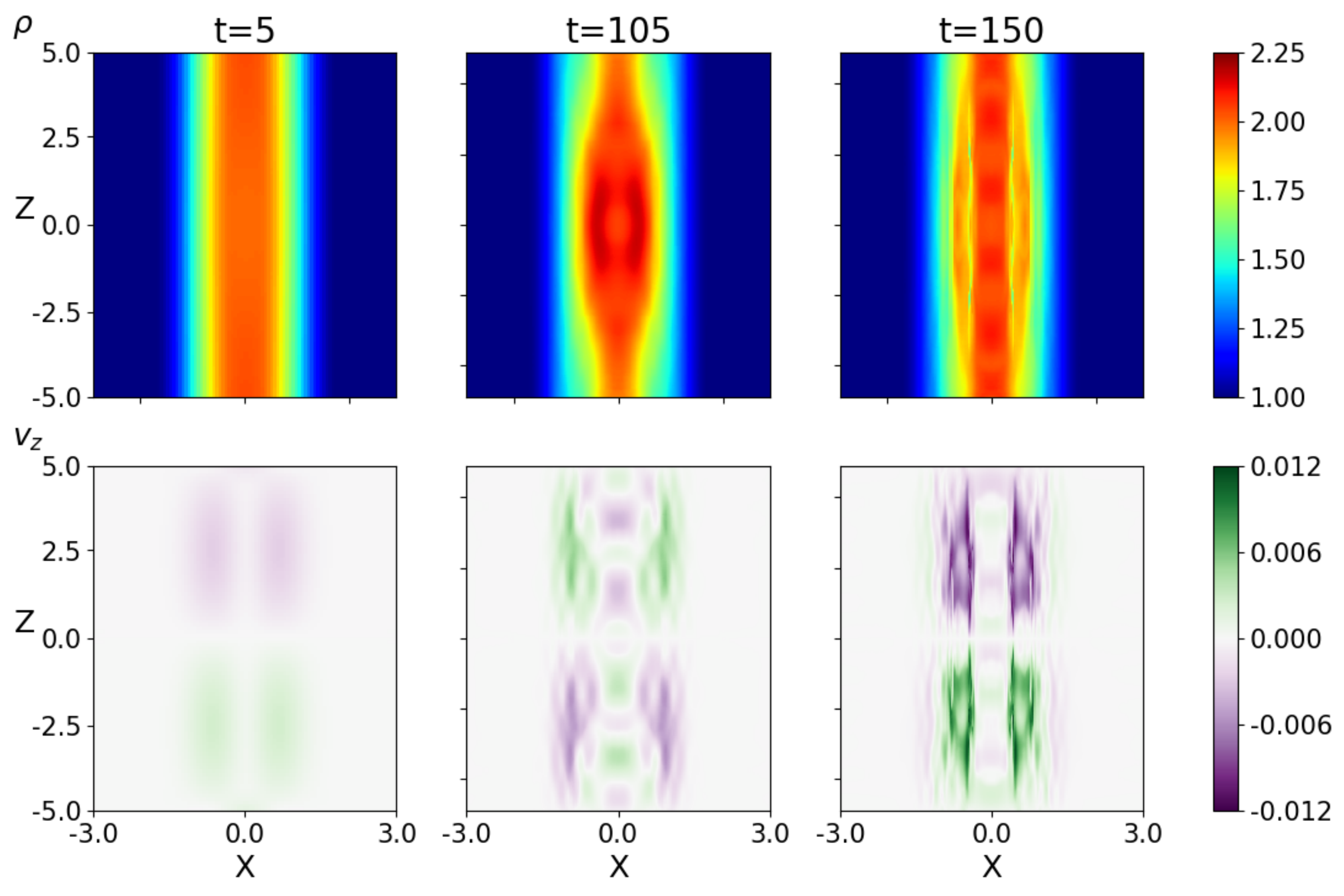}
\caption{Longitudinal cuts of density (top) and longitudinal component of velocity (bottom) at $y=0$ for the thick-layer case at the same simulation times as in Fig.~\ref{denscutthick}. The full temporal evolution is available as an online movie.}
\label{vzthick}
\end{figure*}

The cross-sectional cut of density initially shows no significant density variations, which is consistent with the fact that torsional Alfv\'{e}n waves are incompressible in the linear regime. A  slight periodic compression and expansion of the tube area can be seen, owing to the ponderomotive force as the analytical second-order perturbation predicts \citep[see also][]{Hollweg71}. A better insight into the effect of the ponderomotive force can be visualized in Fig.~\ref{vzthick}. There, we show the temporal evolution  of density and the longitudinal component of velocity in a longitudinal cut along the loop for the thick-layer case. The periodic density enhancement around the tube center caused by the ponderomotive force is evident. The wavelength of the periodic longitudinal flows is half that of the torsional Alfv\'{e}n wave and their amplitude depends quadratically on the amplitude of the azimuthal perturbation. The longitudinal component of velocity changes sign at the center of the tube, which is a converging/diverging point. These results can be compared to those of \citet{Shestov17}  in the case of  weak nonlinear propagating torsional Alfv\'{e}n waves. In their case, the average of the longitudinal component of velocity along the loop is positive. However, since the present simulations correspond to the fundamental mode of standing waves, the longitudinally averaged velocity, $v_z,$ in our case is  zero. We note that the longitudinal velocity develops shear in the nonuniform layer when phase mixing evolves. As is consistent with analytic theory, these longitudinal shear flows are much slower than the corresponding azimuthal flows.

Later in the evolution, the system undergoes a change from a linear to nonlinear regime. The KHi is triggered by the azimuthal shear flows and develops  in the nonuniform layer, as the middle and the right panels of Figs.~\ref{denscutthin} and \ref{denscutthick} evidence. No KHi associated with the longitudinal shear flows is seen in Fig.~\ref{vzthick}. Consistently, the KHi develops earlier and faster in the thin-layer case because of the larger phase-mixing-driven shear flows. Similarly to the results of \citet{Guo19} for their torsional Alfv\'{e}n wave model, we observe that the KHi is first triggered around the middle of the transition region, that is, at $r\approx R$, where the strongest shear flows occur. Hence, the onset of the KHi appears to be a local phenomenon that does not affect, initially, the entire nonuniform region. In connection to this, we interestingly notice  in the bottom right panel of Fig.~\ref{denscutthick}, corresponding to the thick-layer case, that  the KHi is only developing at that instant in a relatively small part of the whole nonuniform layer, while the linear  phase mixing continues to operate in the remaining transitional region.

The most obvious consequence of the onset of the KHi is the formation of eddies clearly seen in the evolution of density. The internal and external plasmas mix as a result of these vortical motions. As time increases, eddies break down to form smaller and smaller structures. The instability drives the flux tube to a turbulent state. The extent of the turbulent zone increases with time and surpasses the width of the nonuniform layer. At sufficiently long times beyond those simulated here, the whole tube should become turbulent (see \citealt{Karampelas2018} for simulations of kink waves). The simulations  show that the turbulence develops in the transverse plane to the magnetic field only. Figure~\ref{vzthick} evidences that there is no eddy formation in the longitudinal direction. Thus, we are in a clear situation of nearly 2D Alfv\'enic turbulence in which the spatial scales perpendicular to the background magnetic field are much smaller than the scales in the magnetic field direction. As shown in Sect.~\ref{sec:turbulence}, this has implications for the energy cascade scaling law.

Since we are studying the fundamental standing mode, the amplitude of the torsional oscillations is maximal at the middle of the tube ($z=0$) and zero at the tube ends ($z=\pm L/2$) because of the line-tying condition. Consequently, it is at the $z=0$ plane that turbulence develops fastest and strongest. Conversely, at $z=\pm L/2$ the KHi is not triggered, so there is no development of turbulence. 

The nonlinear evolution of the torsional oscillations shares many similarities with that of kink MHD modes \citep[see, e.g.,][]{Terradas08,Antolin15,Magyar16,Howson17,Terradas18,Karampelas19,Antolin19}. In our case, the tube is not displaced laterally as it is for a kink mode, but the KHi develops in a similar fashion. In the case of the kink mode in a transversely nonuniform tube, a previous step is the energy transfer from the global lateral oscillation to localized Alfv\'en modes in the nonuniform layer owing to resonant absorption. \citep[see, e.g.,][]{terradas06,goossens2011,arregui2011,Soler15}. These localized Alfv\'en modes with kink azimuthal symmetry are the ones that phase mix and eventually trigger the KHi. We note that the KHi can also be triggered by the kink mode even in tubes with an abrupt density transition \citep[see][]{Antolin19}. In that case, resonant absorption does not happen initially, and it is the own azimuthal shear associated with the global kink mode perturbations that triggers the KHi. For the torsional oscillations studied here, there is no global mode, and the resonant absorption mechanism does not occur. In our case, Alfv\'en modes with torsional azimuthal symmetry are already excited by the initial condition, so the phase mixing starts to operate from the beginning of the simulation.

\begin{figure*}[htbp]
\centering
\includegraphics[width=1.85\columnwidth]{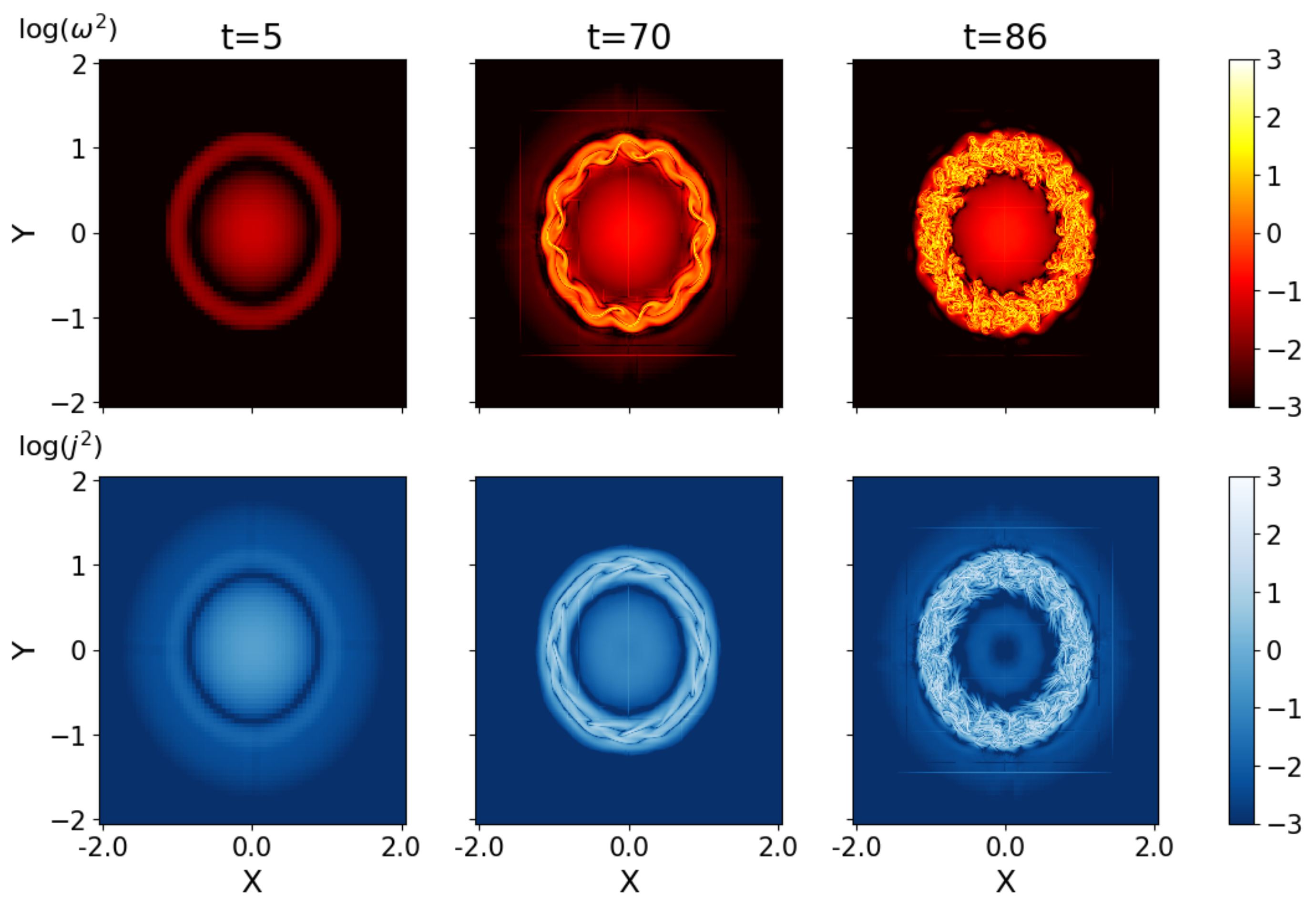}
\caption{\textit{Top}: Cross-sectional cut at $z=0 $ of  vorticity  squared (in logarithmic scale) for the thin-layer case at three different simulation times indicated above each panel. \textit{Bottom}: Same, but for the current density squared at $ z \approx L/2$. The straight lines seen in some panels are visualization artifacts at the boundaries of different AMR patches. These artifacts are not present in the actual simulation data. The full temporal evolution is available as an online movie.}
\label{curvortcutthin}
\end{figure*}

\begin{figure*}
\centering
\includegraphics[width=1.85\columnwidth]{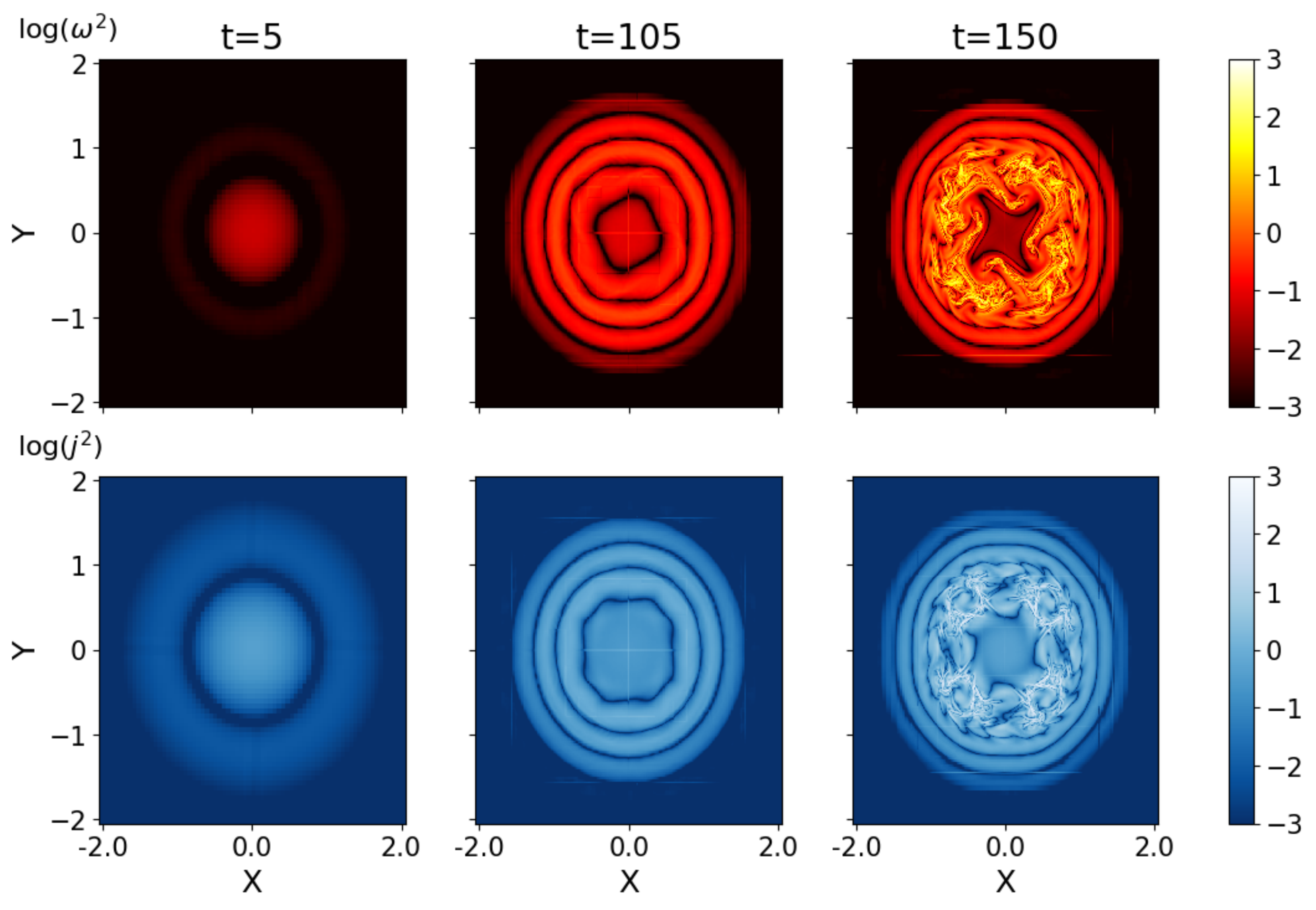}
\caption{Same as Fig. \ref{curvortcutthin}, but for the thick-layer case. We note that the three selected simulation times are different from those of Fig.~\ref{curvortcutthin}. The full temporal evolution is available as an online movie.}
\label{curvortcutthick}
\end{figure*}

\subsection{Increase of vorticity and current density}
\label{sec:vort}

Two parameters that help us illustrate the important effect of the KHi on the torsional oscillation evolution are vorticity, $ \mbox{\boldmath{$\omega$}}  = \nabla \times \mathbf{v}$, and current density, $\mathbf{j} = \mu^{-1}\nabla \times \mathbf{B}$. We find that the KHi dramatically increases both quantities because of the small scales that rapidly show up once the KHi is triggered. We compute the vorticity squared and the current density squared in a cross-sectional cut of the tube. The cut is done at the tube center, $z=0$, for the vorticity and near one tube end, $z \approx L/2,$ for the current density. The reason for choosing these different cuts is the different spatial dependences of the two parameters along the tube: vorticity is maximal at the tube center and zero at the ends, while the opposite applies to current density. Figures~\ref{curvortcutthin} and \ref{curvortcutthick} show these results for the thin-layer and thick-layer cases, respectively. We used logarithmic scale to better visualize the important change in the magnitude of both quantities.

\begin{figure}[htbp]
\centering
\resizebox{\hsize}{!}{\includegraphics{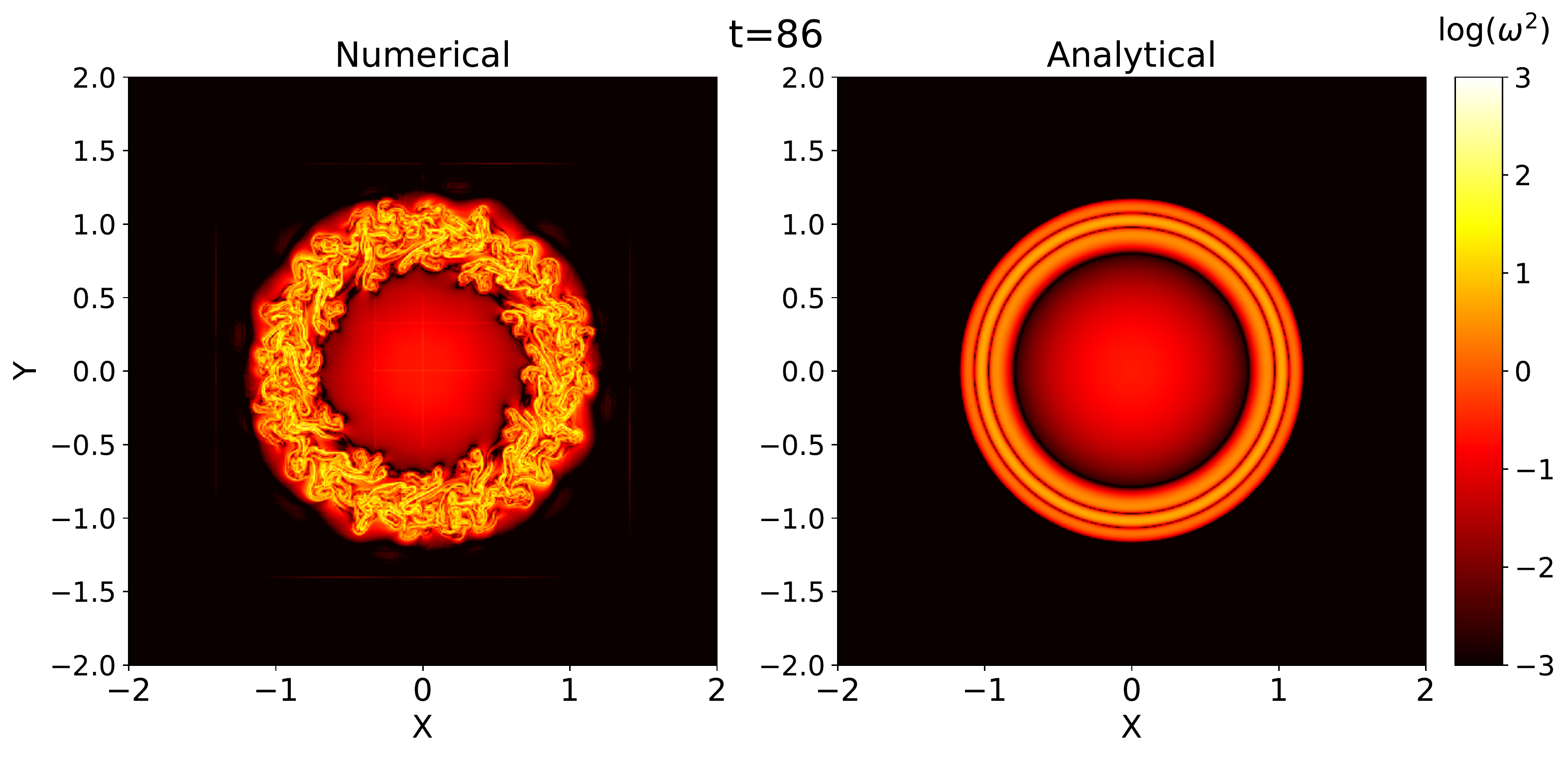}} 
\resizebox{\hsize}{!}{\includegraphics{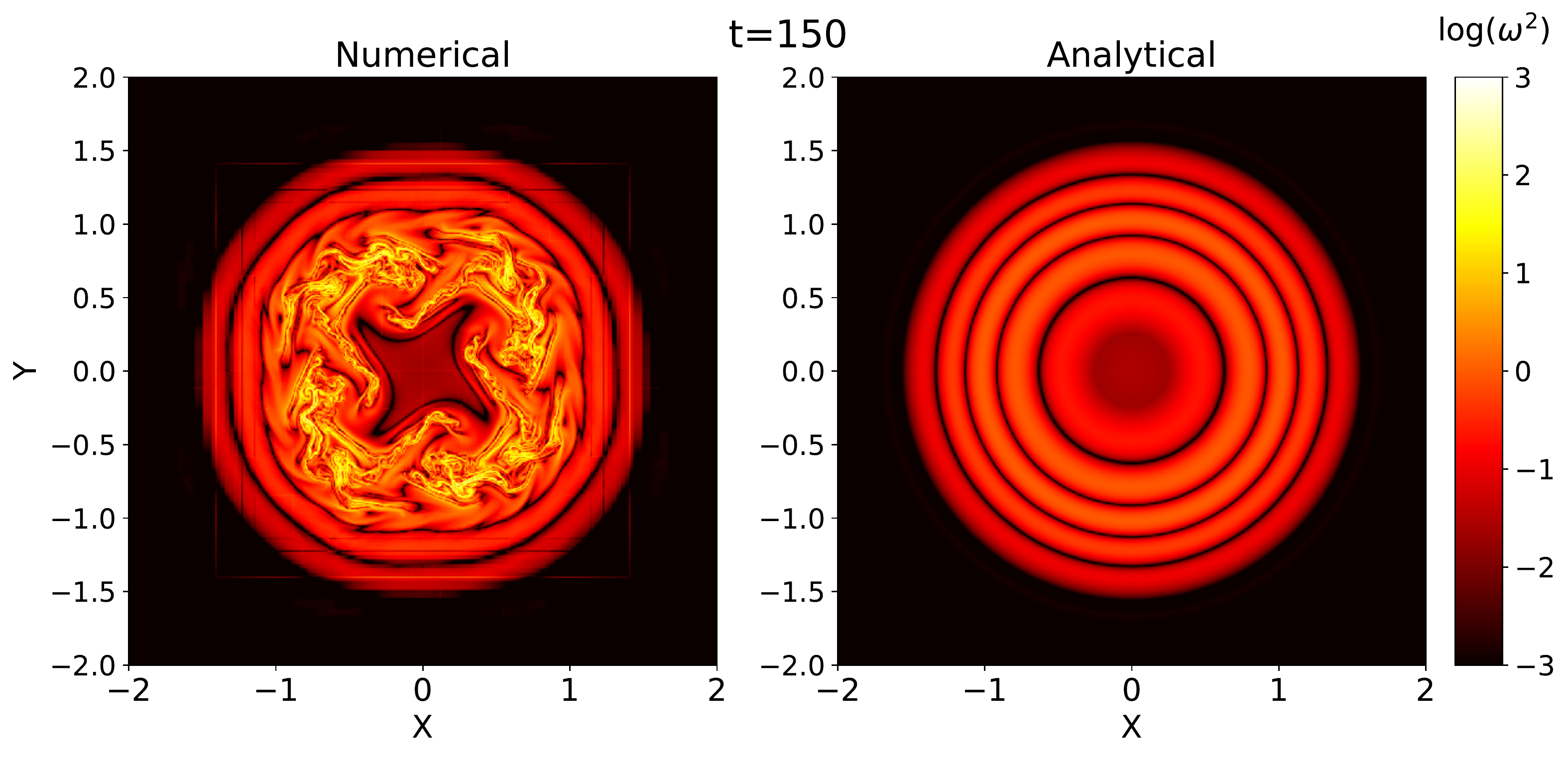}} 
\caption{\textit{Top}: Cross-sectional cut at $z=0 $ of  vorticity  squared (in logarithmic scale) for the thin-layer case at the final simulation time. The left and right panels correspond to the nonlinear numerical results and the linear analytic results, respectively. \textit{Bottom}: Same,  but for the thick-layer case.}
\label{vortsimnumthin}
\end{figure}

As expected, both vorticity and current density initially agree with the results from linear analytic theory. Thus, they  evolve in the nonuniform layer, and their values increase as a result of phase mixing alone. However, once the KHi sets in, analytic, and numerical results start to differ substantially. The full nonlinear evolution is characterized by the generation of very fine structures in both vorticity and current density, and by a dramatic increase of their magnitudes. Figure~\ref{vortsimnumthin} shows a comparison between linear analytic results and nonlinear numerical results for the vorticity squared at the final frame of the simulations. The analytic vorticity is computed with  the azimuthal velocity component  given in Eq.~(\ref{vphi}).  The comparison in the case of the current density squared reveals similar results and for the sake of simplicity is not shown here. The very fine structures in vorticity and current density associated with the KHi can also be seen in previous works of nonlinear kink oscillations \citep[see e.g.,][]{Antolin14,Howson17,Antolin19} where the spatial distribution is slightly different because of the different azimuthal symmetry of the perturbations.

\begin{figure}[htbp]
\centering
\includegraphics[width=\columnwidth]{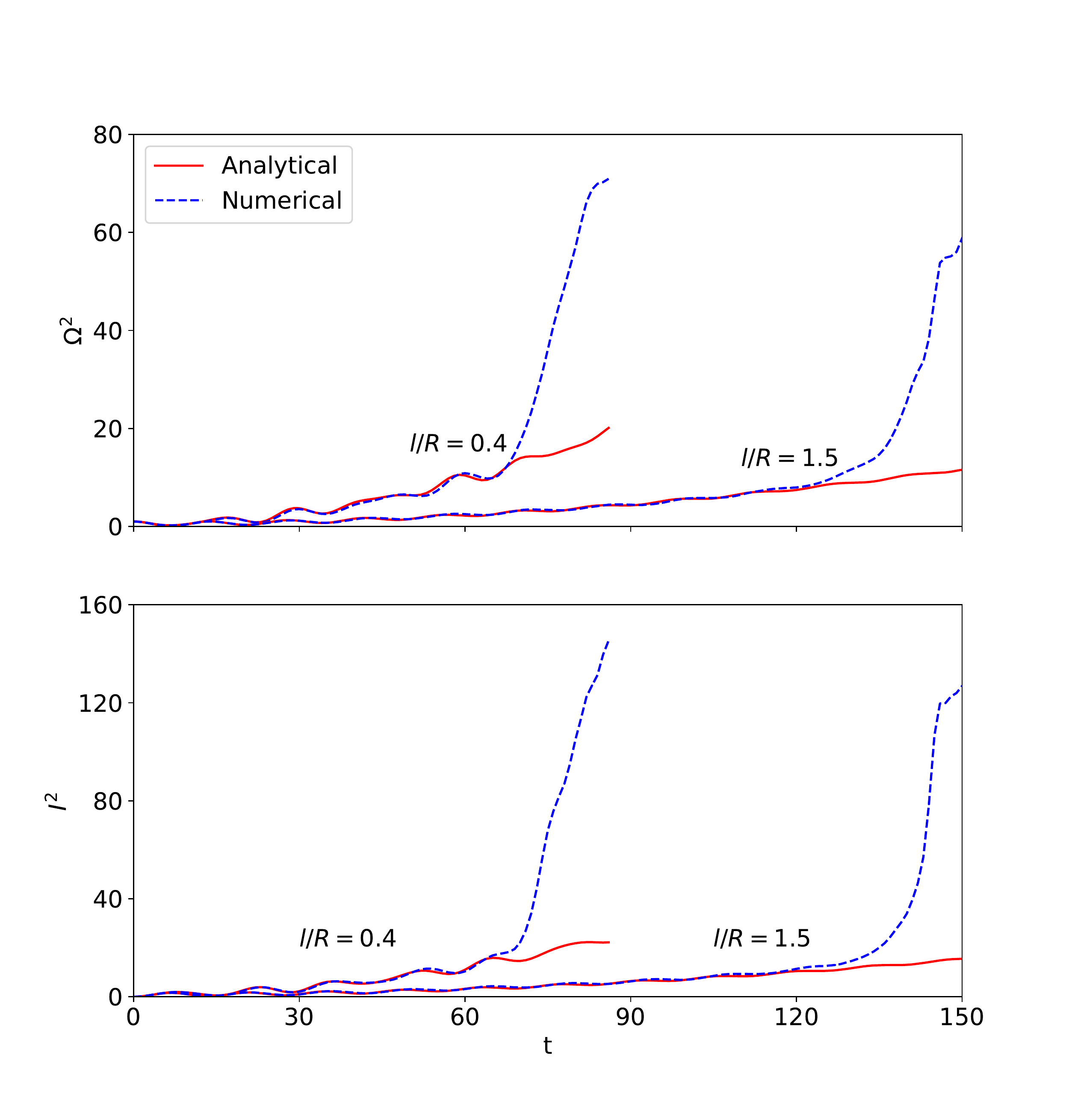}
\caption{Vorticity squared (\textit{top panel}) and current density squared (\textit{bottom panel}) integrated over the whole computational domain as a function of simulation time for both thin-layer ($ l/R=0.4 $) and thick-layer ($ l/R=1.5 $) cases. Red solid lines correspond to linear analytic results, and blue dashed lines correspond to  nonlinear numerical results.  Vorticity  is normalized to the value at $t=0$. No normalization is needed for current density  since it vanishes at $t=0$.}
\label{curvortcuttot}
\end{figure}

The change of magnitude of vorticity squared and current density squared  can be better studied by integrating both quantities over the whole computational domain as
\begin{eqnarray}
\Omega^{2}(t) &=& \int \omega^{2}(\mathbf{r},t) d^{3}\mathbf{r},\label{vortsq} \\
I^{2}(t) &=& \int \mathbf{j}^{2}(\mathbf{r},t) d^{3}\mathbf{r}.\label{cusq} 
\end{eqnarray}
Figure~\ref{curvortcuttot} shows the evolution of $\Omega^{2}$ and $I^{2}$ for both thin-layer and thick-layer cases. A comparison with the values predicted by linear theory is also included. 

We find that both integrated quantities increase with time following a slightly oscillatory pattern, and numerical and analytic results initially agree. The increase is faster in the thin-layer case than in the thick-layer case. Such a result is consistent with the behavior of phase mixing \citep[see, e.g.,][]{HeyvaertPriest83}. Hence, the initial increase of vorticity and current density is predicted by linear theory and is caused by phase mixing alone. However, once the KHi is triggered in the numerical simulations, both integrated quantities suffer a dramatic increase that is not predicted by linear theory. At the end of the simulations, the numerical values of $\Omega^{2}$ and $I^{2}$ are significantly larger than those of linear theory.

Previous works have studied  vorticity in the case of nonlinear kink waves \citep[e.g.,][]{Terradas08,Antolin15,Karampelas17,Howson20}. \citet{Guo19} studied the average $z$-component of vorticity squared integrated at the loop apex for both torsional Alfv\'{e}n and kink waves. As in these works, we verified that the main contribution to the vorticity squared is its $ z$-component due to the development of strong gradients in the transverse components of velocity. \citet{Guo19} and \citet{Howson20} found increasing oscillations in vorticity with time not only due to phase mixing and the KHi, but also because of the continuous driver used to excite the waves. Conversely, here we impose an initial perturbation and let the system evolve. Therefore, in our case, the generation of vorticity and current density is because of phase mixing in the linear regime and the KHi in the nonlinear regime, with no influence from an external driver.

At this point, we feel it is relevant to stress the importance of using a sufficiently high resolution to correctly capture the very fine scales generated because of the KHi. For instance, Fig. 2 of \citet{Antolin15} shows a cross-sectional cut of the $z$-component of the vorticity and current density at the apex of a cylindrical tube representing a prominence thread oscillating in the kink mode. They show the same results with low resolution (left panels) and high resolution (right panels). Their high-resolution simulation is able to describe smaller scales and shows higher values of current density and  vorticity. Another example can be found in \citet{Howson17}, where they compare ideal MHD simulations of nonlinear kink oscillations with simulations including viscosity and/or resistivity. Their Fig.~6 shows that in the ideal MHD simulation, vorticity increases drastically due to the development of KHi, but after $t\approx 700$~s it saturates and decreases, presumably owing to numerical dissipation, which also explains the loss of kinetic energy for $t> 700$~s (see their Fig. 9). In another ideal simulation but at lower resolution, they show  that the saturation of vorticity happens earlier. Simulations including physical dissipation show similar results, pointing out that physical and numerical dissipation can play equivalent roles. If dissipation is strong enough, the KHi can be delayed or can even be suppressed altogether.  In the case of torsional Alfv\'{e}n waves, \citet{Guo19}, in their Fig. 2, did not find the significant increase of vorticity that we obtained after the onset of the KHi. Possible explanations may be that the KHi had not yet completely developed in their simulation, or that  numerical diffusion plays a role. 

\begin{figure}[htbp]
\centering
\resizebox{\hsize}{!}{\includegraphics{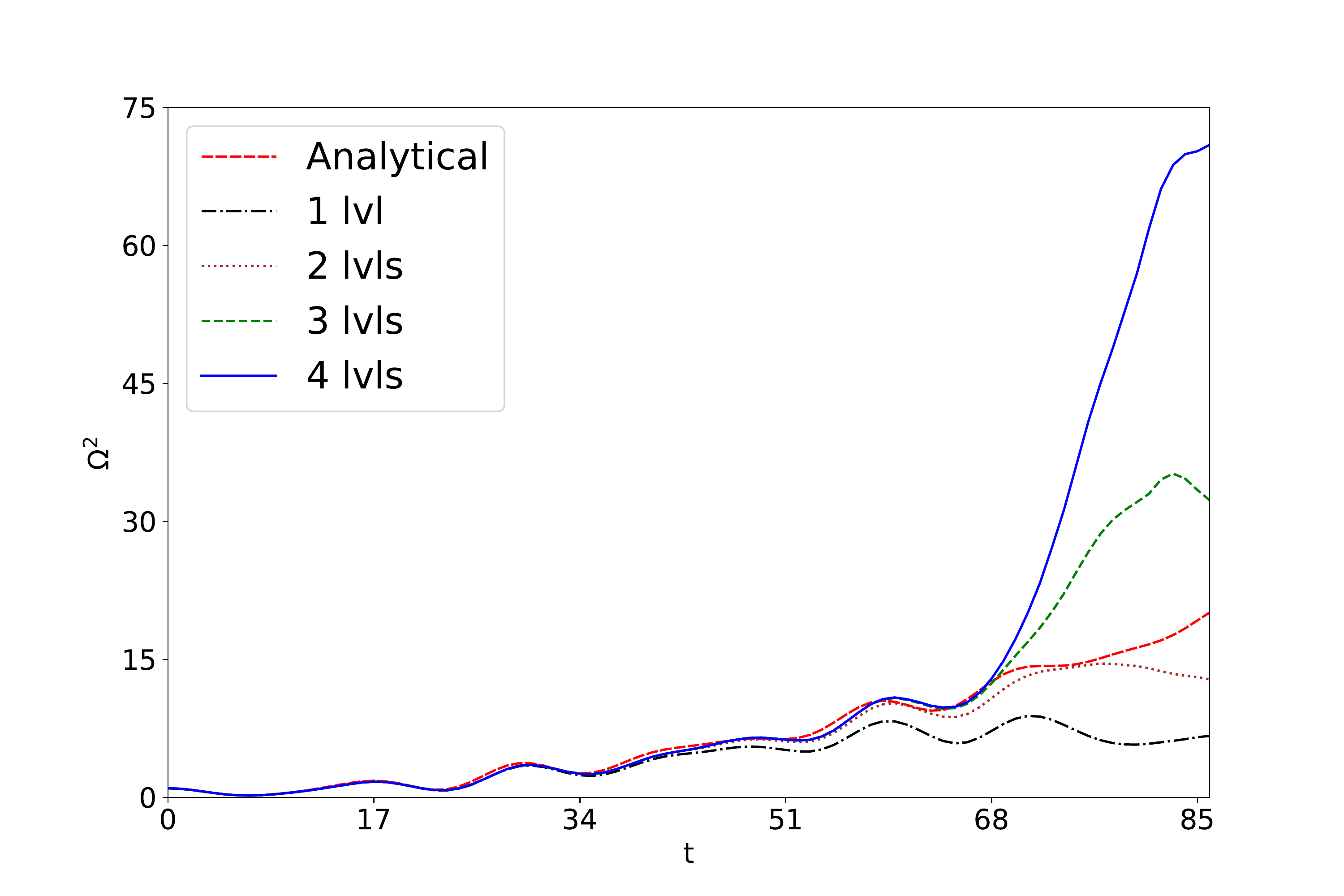}}
\caption{Vorticity squared integrated over the whole computational domain as a function of simulation time for the thin-layer case. The various lines correspond to simulations with different levels of refinement in the AMR scheme: 1 (black dot-dashed line), 2 (brown dotted line), 3 (green dashed line), and 4 (blue solid line). For comparison purposes, the linear analytical value is also shown (red dashed line).  Vorticity  is normalized to the value at $t=0$.}
\label{resolution}
\end{figure}

To further investigate the influence of numerical resolution, we have repeated the simulation for the thin-layer case, but using fewer levels of refinement in the AMR scheme. The lower the number of levels, the lower the effective resolution. In Fig. \ref{resolution}, we show the evolution with time of vorticity squared integrated over the whole computational box for these test simulations.  These results are also compared to those of  linear theory. In all cases and for comparison purposes, the end time of the simulations has been set to that of the four-level simulation. We find that all simulations behave similarly in the linear regime, but after the system transits to the nonlinear regime and the KHi is triggered, the higher the resolution, the larger the increase of  vorticity. We obtain the interesting result that the simulations with only one or two levels of refinement are not even able to correctly recover the linear results. In those cases, numerical dissipation is also strong enough to prevent the onset of the KHi, in agreement with \citet{Howson17}. These results confirm the need to use high resolutions to correctly describe the evolution of the KHi and to minimize the significant impact of numerical diffusion.

\subsection{Investigating the onset of the KHi}

The aim here is to compare the onset time of the KHi obtained from the simulations with the theoretical one predicted by \citet{Browning}. The problem with the simulations is to find an objective criterion to define the onset time. The behavior with time of the integrated vorticity squared   can be a way to estimate the onset time from the simulation data. As seen in Fig.~\ref{curvortcuttot}, the slope of the ascending trend of  vorticity changes and deviates from the linear result at a specific time that can be associated with the onset of the KHi. We denote this time as $\tau_{\rm KH}$. The integrated current density squared also displays a slope change at the same time as vorticity, approximately, but we used the vorticity data. Then, by visual inspection, we determined that $\tau_{\rm KH} \approx 66$ in the thin-layer case, and $\tau_{\rm KH} \approx 120$ in the thick-layer case. Moreover, we compared these estimated onset times with Eq.~(\ref{tcri}) obtained by \citet{Browning} in the strong phase-mixing limit. Equation~(\ref{tcri}) gives $t_{\rm crit} \approx 20$ in the thin-layer case and $t_{\rm crit} \approx 75$ in the thick-layer case. Therefore, Eq.~(\ref{tcri}) predicts that the KHi is triggered  earlier than the simulations apparently show.

A reason for this discrepancy may lie  in the strong phase-mixing approximation behind Eq.~(\ref{tcri}). When this approximation is relaxed, we can use the following method provided in Sect. 4 of \citet{Browning} to adapt
Eq.~(\ref{tcri}) to the weak phase-mixing case:~\begin{equation}
t_{\rm crit} =T_{\rm crit}\left( \Omega_{1} \right)\frac{l}{A(R) v_0},
\label{tcri2}
\end{equation}
where $T_{\rm crit}$ is their dimensionless onset time, which depends on the parameter $\Omega_{1}$ that, in our notation, is
\begin{equation}
\Omega_{1}=\frac{k_{z} l}{\varepsilon},
\label{omega1}
\end{equation}
with $ k_{z}  = \pi/L$ the parallel wave number to the magnetic field. Figure~7 of \citet{Browning} shows the dependence of $T_{\rm crit}$ with $\Omega_{1}$. The strong phase-mixing limit should correspond to the case $\Omega_{1} \gg 1$ so that $T_{\rm crit} \to 4$ and Eq.~(\ref{tcri}) is recovered. We note that in their Sect.~4, \citet{Browning} considered the particular case $v_{\rm A,i} = 0.5 v_{\rm A,e}$, while in our model $v_{\rm A,i} =  v_{\rm A,e} / \sqrt{2} \approx 0.71 v_{\rm A,e}$. However, we assume that the results in their Fig.~7 remain approximately valid in our case. According to the parameters of our simulations, we have $ \Omega_{1}= 2\pi/5 \approx 1.26$ for the thin-layer case and  $ \Omega_{1}= 3\pi/2 \approx 4.71 $ for the thick-layer case. Therefore, we are far from the strong phase-mixing limit, especially in the thin-layer case. Using Fig.~7 of \citet{Browning}, we approximately determine that $T_{\rm crit} \approx 8$ for the thin-layer case, so $t_{\rm crit} \approx 40$ according to the modified Eq.~(\ref{tcri2}). This critical time approaches the value inferred from the integrated vorticity slope change but is still lower. For the thick-layer case, the determination of $T_{\rm crit}$ is more problematic since Fig.~7 of \citet{Browning} stops at $\Omega_{1} =2$ where $T_{\rm crit} \approx 7$. As \citet{Browning} explained, we can assume that $T_{\rm crit}$ approaches the strong phase-mixing value, $T_{\rm crit} = 4, $ asymptotically when $\Omega_{1} \gg 1$, but for $ \Omega_{1}\approx 4.71 $ we are probably still far from the limit. An educated guess would be to assume that $T_{\rm crit}$ is somewhere between the asymptotic value and the last value seen in Fig.~7 of \citet{Browning}. So, we roughly approximate $T_{\rm crit} \approx 5$, which results in $t_{\rm crit} \approx 94$ in the thick-layer case according to the modified Eq.~(\ref{tcri2}). This value is again lower than that obtained from the vorticity data.

One may ask why the critical times  of \citet{Browning} are systematically smaller than those inferred from the slope change of the integrated vorticity squared, even when the strong phase-mixing limit is relaxed. The answer to this question is that $t_{\rm crit}$ of \citet{Browning} corresponds to the time at which the KHi is {\em \emph{locally}} excited within the nonuniform later, while our $\tau_{\rm KH}$ should be associated with a time for which the KHi has developed enough to  {\em \emph{globally}} impact the flux tube dynamics. We recall that $\tau_{\rm KH}$ is obtained from a quantity that has been integrated over the whole tube, so that local effects are probably blurred.  To check this statement, we need a method to estimate, from the simulations, the time at which the KHi is first locally triggered in the nonuniform layer. The evolution of density (see again Figures~\ref{denscutthin} and \ref{denscutthick}) indeed shows that the initial local distortion of density associated with the KHi happens for a somewhat earlier time than $\tau_{\rm KH}$, but we need a more robust approach. 

Inspired by \citeauthor{Terradas18} (\citeyear{Terradas18}; see also \citet{Antolin19}),we studied the excitation of different azimuthal wave numbers using the azimuthal and radial components of velocity in the transitional layer. We considered a cross-sectional cut at the center of the tube, $z=0$, and extracted, from the simulations, the values of the azimuthal and radial components of velocity at the middle of the transition region, $r=R$, as functions of the azimuthal angle from $ 0$ to $2\pi$. After that, we applied the discrete Fourier transform to the data using the fast Fourier transform (FFT) algorithm  in 1D \citep{Cooley65}. Following the notation of \citet{Terradas18}, the discrete Fourier transform can be defined as
\begin{equation}
G(p)=\sum^{N-1}_{k=0} g(k) \exp{\left(-\frac{2\pi ipk}{N}\right)},
\label{fftdiscraz}
\end{equation}
where  $N$ is the number of samples (points), $g(k)$ is the angular sampling of the azimuthal/radial velocity, and $p = 0, \dots , N-1$ is an integer that plays the role of the azimuthal wave number. $p=0$ is the torsional or sausage mode, $p=1$ is the kink mode, and $p \ge 2$ are the fluting modes \citep[see, e.g.,][]{Edwin83}. We find that the Fourier coefficients, $G(p)$, are complex except for $p=0$, which is always real. \citet{Terradas18}  found that the Fourier coefficients associated with the azimuthal component of velocity were purely imaginary, while those associated with the radial component were real. They explained that this result is due to parity rules of their initial condition since they only excite the $p=1$ kink mode at $t=0$. Nonetheless, we excited the $p=0$ torsional mode at $t=0$, and those parity rules may not  apply in our case.

\begin{figure}[!tbp]
\centering
\resizebox{\hsize}{!}{\includegraphics{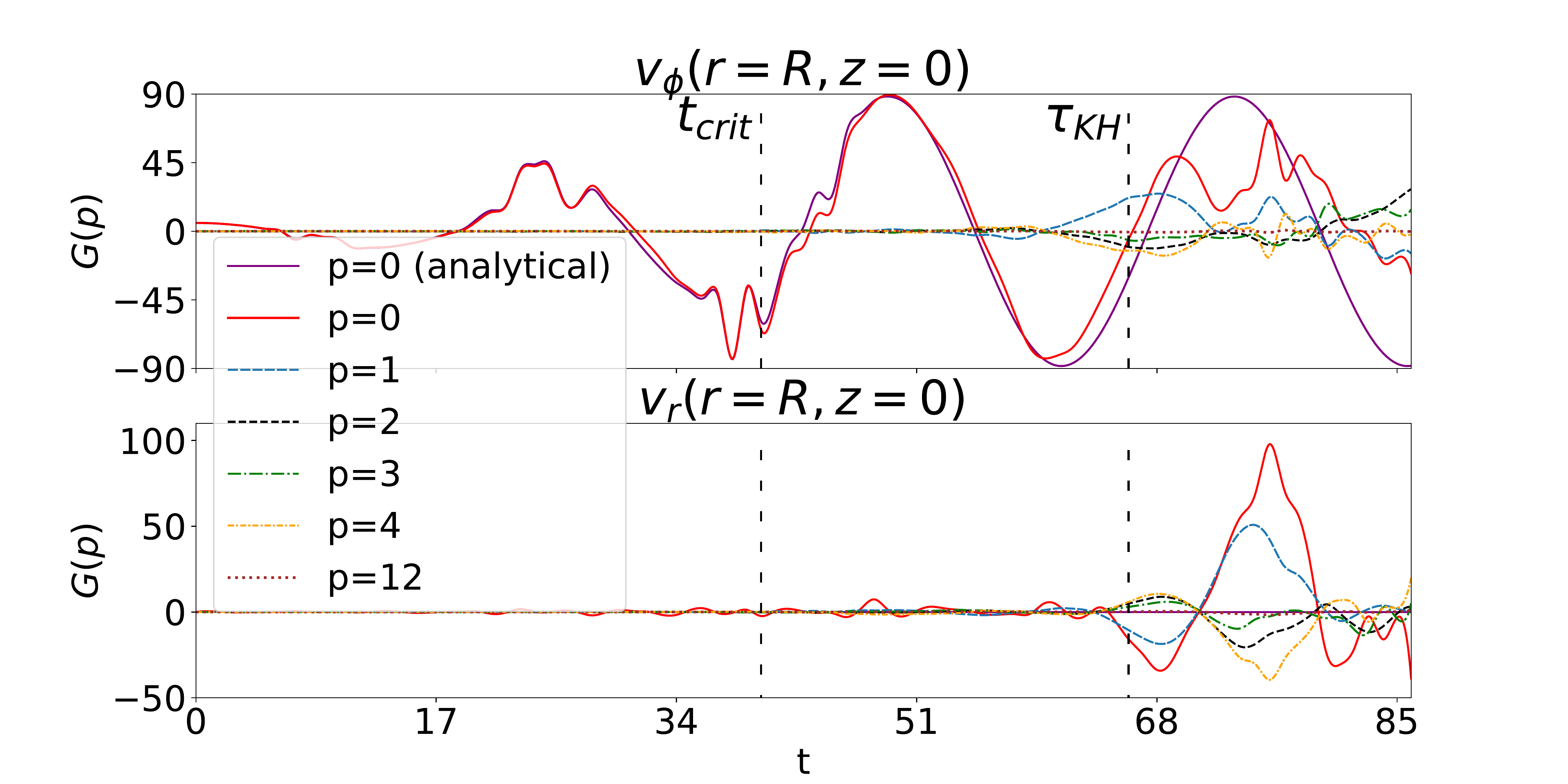}}
\caption{\textit{Top panel}: Real part of the Fourier coefficients with $p=$~0, 1, 2, 3, 4, and 12 as functions of the computational time  for an angular sampling of the azimuthal component of velocity at $r=R$ and $z=0$ in the thin-layer case. The two vertical dashed lines correspond to the critical time of \citet{Browning}, $ t_{\rm crit}= 40$, and the onset time inferred from the integrated vorticity slope change, $ \tau_{\rm KH}= 66$. The linear analytic result, where only the $p=0$ mode is present, is overplotted for comparison.   \textit{Bottom panel}: Same as top panel but for the radial component of velocity.}
\label{azimthin}
\end{figure}

\begin{figure}[!tbp]
\centering
\resizebox{\hsize}{!}{\includegraphics{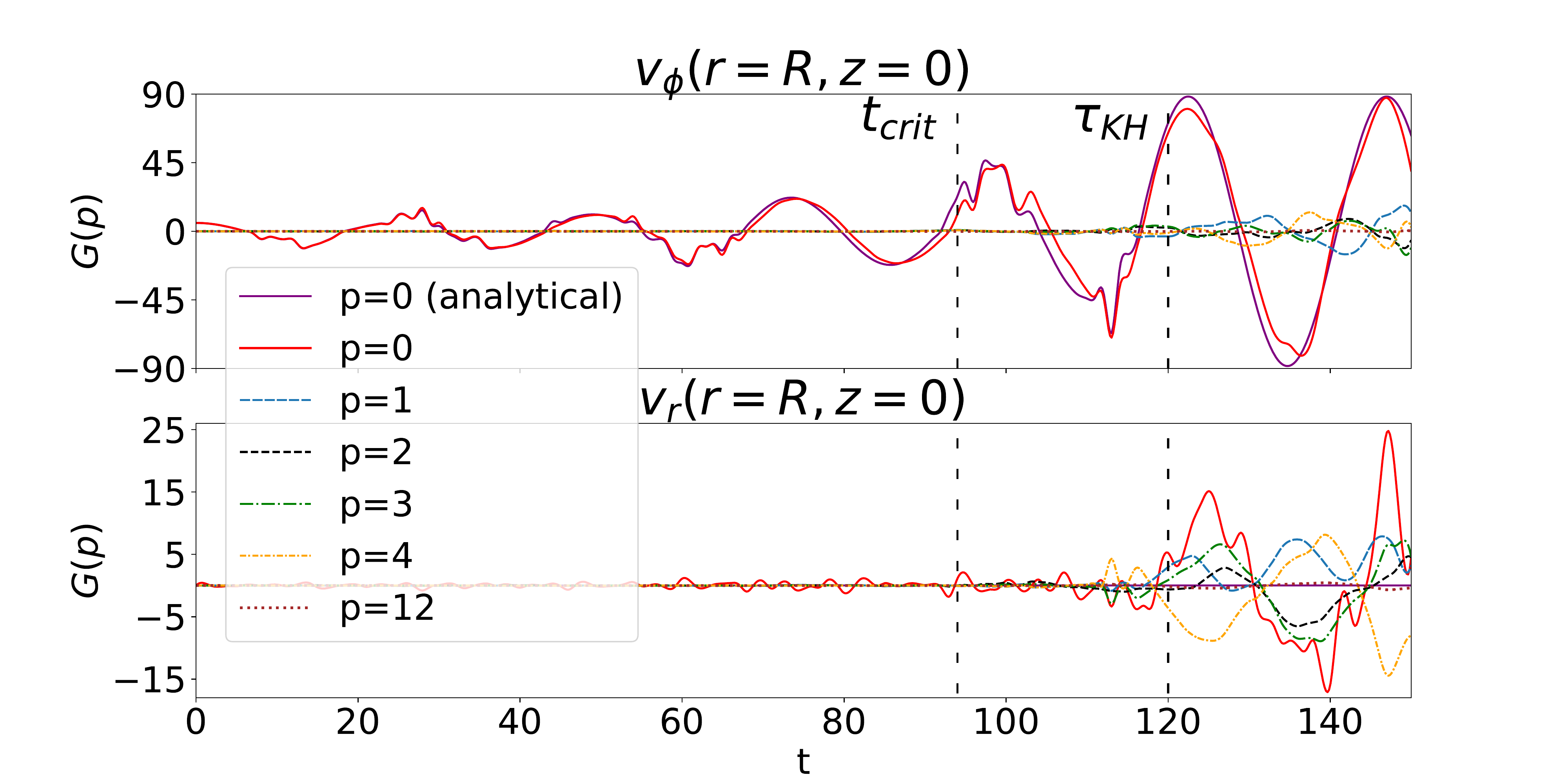}}
\caption{Same as Fig.~\ref{azimthin}, but for the thick-layer case. Here, $ t_{\rm crit}= 94$ and $ \tau_{\rm KH}= 120$.}
\label{azimthick}
\end{figure}

The results of the azimuthal Fourier analysis are shown in Fig.~\ref{azimthin} for the thin-layer case and in  Fig.~\ref{azimthick} for the thick-layer case. In both figures, the top (bottom) panels show the real part of the Fourier coefficients as functions of time obtained from the azimuthal (radial) component of velocity. The studied azimuthal wave numbers are $p=0$ (torsional), $p=1$ (kink), $p=2$ (first fluting), $p=3$ (second fluting), $p=4 $ (third fluting), and $ p=12$ (eleventh fluting). Additionally, we also plot the result predicted by linear theory, which only contains the $p=0$ torsional mode. With vertical dashed lines, we indicate the values of $t_{\rm crit}$ and $\tau_{\rm KH}$ for each case.

Concerning the azimuthal component of velocity, the $p=0$ Fourier coefficient is dominant during the linear regime, while the others are negligible. This remains the case until the development of KHi, which excites higher azimuthal wave numbers. This agrees with the results of \citet{Terradas18} and \citet{Antolin19}. We obtain an amplification of the $p=0$ Fourier coefficient as time increases. This amplification is due to phase mixing alone,  as can be seen by  comparing the numerical result with that from linear theory. After the KHi is triggered, not all azimuthal wave numbers are equally excited at a given time. The larger the value of $p$, the later these modes are excited.   Toward the end of the simulations and especially in the thin-layer case, we observe  that the  amplitude of the Fourier modes with high azimuthal wave numbers can even become comparable to that of the torsional mode because of the nonlinear evolution of the KHi into turbulence. To fully understand why some azimuthal wave numbers are excited before others, a similar theoretical study to that of \citet{Browning}  but in cylindrical geometry and with time-varying flows is required. This is beyond the scope of the present paper.

We verified (not shown here) that the excitation of high-order azimuthal wave numbers initially occurs  in the nonuniform region only. There is no signature of azimuthal wave numbers other than $p=0$ in the uniform core of the tube  where the triggering of the KHi is not possible. However, once the KHi is fully developed and the whole tube is driven to a turbulent state, high-order azimuthal wave numbers can be present everywhere.

Regarding the radial component of velocity, we consistently see that all Fourier modes vanish at $t=0$ and remain negligible in the linear regime before the excitation of the KHi, since the oscillations are strictly polarized in the azimuthal direction at the beginning. After the triggering of the KHi, nonzero amplitudes of $G(p)$ for the radial component of velocity are found, with the $p=0$ mode being the largest contributor.

Recovering the results of the density evolution shown in Figs.~\ref{denscutthin} and \ref{denscutthick}, one could tentatively infer that the dominant azimuthal wave numbers are $p=12$ and $p=4$ in the thin-layer and thick-layer cases, respectively, merely by counting the number of big vortices that appear when the KHi develops. This simple estimation can be compared with the azimuthal Fourier analysis. In Fig.~\ref{azimthick}, we can see that the $p=4$ mode is one of the most relevant modes after the torsional $p=0$ mode in the thick-layer case at the end of the simulation, especially for the radial component of velocity. Conversely, in the thin-layer case (Fig.~\ref{azimthin}), the $p=12$ mode is hardly excited in either velocity component. There are several reasons that may explain this discrepancy. For instance, we note that the analysis of azimuthal modes preformed here does not use the density but the velocity components, and is done at a particular radial position, $r=R$. We checked (not shown here) that the relative amplitude of a particular Fourier mode depends on both the variable used in the analysis and the radial position where the azimuthal sampling is done.

In both thin-layer and thick-layer cases, the critical times of \citet{Browning}, $t_{\rm crit}$, approximately coincide with the times for which azimuthal wave numbers with $p\neq 0$ start to grow within the nonuniform layer. This result confirms us that the onset times obtained from the change of trend in the integrated vorticity squared, $\tau_{\rm KH}$, overestimate the actual {\em \emph{local}} onset of the KHi. Instead, $\tau_{\rm KH}$ should be interpreted as a timescale for which the KHi starts to have a {\em \emph{global}} effect on the flux tube dynamics. In Sect.~\ref{parstudy}, we detail a study of the effect of various model parameters on the value of $\tau_{\rm KH}$.

\subsection{Turbulent energy cascade to small scales}
\label{sec:turbulence}

The results of the numerical simulations clearly indicate that the generation of small spatial scales speeds up when the KHi is triggered, and turbulence develops thereafter. Here, we study how the energy cascades from large to small scales and compare the efficiency of this process in the linear and nonlinear regimes. For this purpose, we calculated the amplitude spectrum of the averaged energy of the perturbations as a function of time.  The procedure is described below.

\begin{figure}[htbp]
\centering
\resizebox{\hsize}{!}{\includegraphics{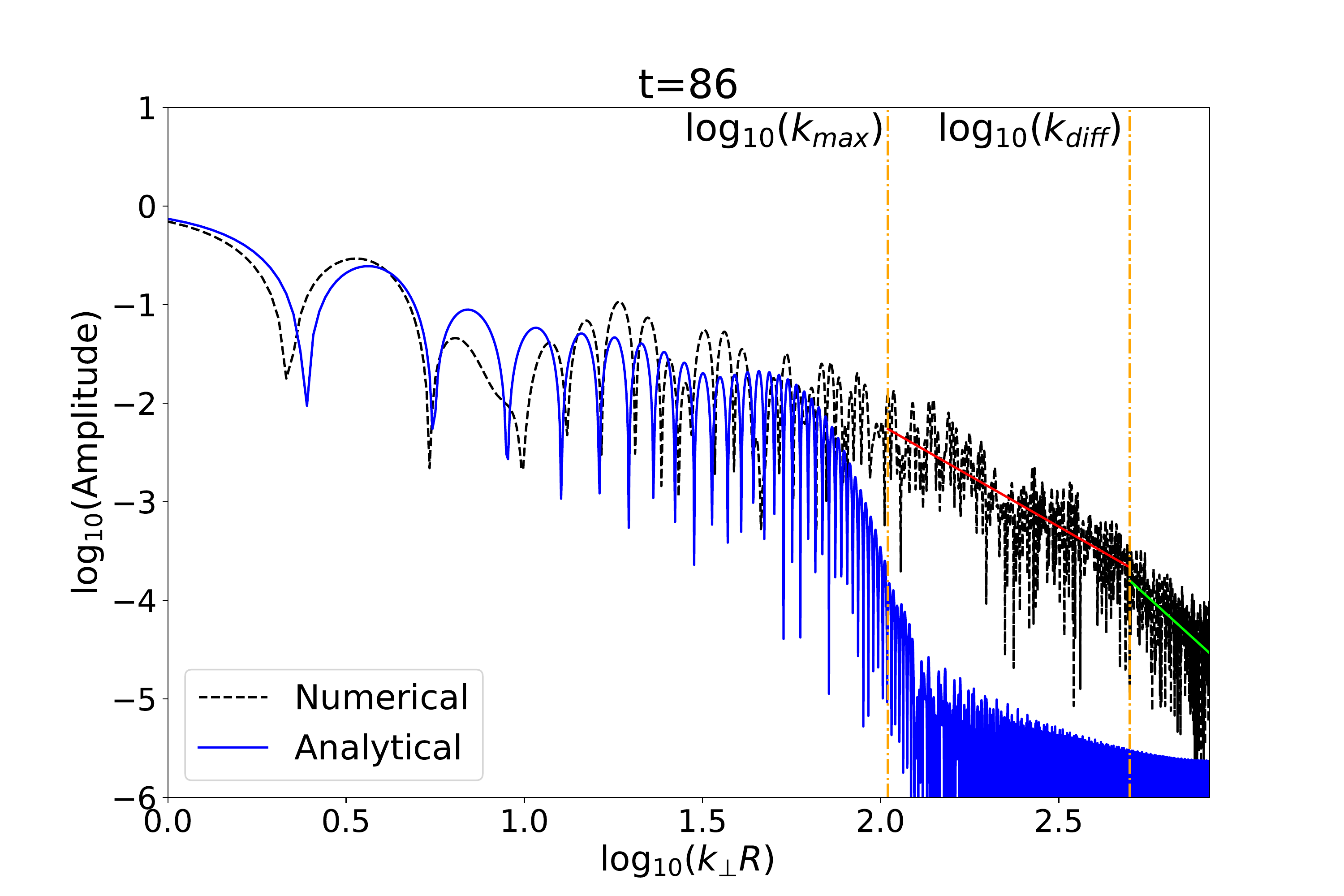}}\caption{Amplitude spectrum (arbitrary units) of the averaged total energy of the perturbations for the thin-layer case at the end of the simulation for both numerical and linear analytic results. The vertical  dot-dashed orange lines denote the maximum phase-mixing-generated wave number, $k_{\rm max} R \approx 105$, obtained from Eq.~(\ref{effwnu}), and an estimated wave number for which numerical diffusion starts to play a role, $k_{\rm diff} R \approx 500$. The red line is a least-squares linear fit for $k_{\rm max} < k_\perp < k_{\rm diff}$ in log-log scale whose slope is $-2.07 \pm 0.02$. The green line is the same fit but for $k_\perp > k_{\rm diff}$ , whose slope is $-3.28 \pm 0.07$}
\label{ffttransporthin}
\end{figure}

First, we compute the sum of the magnetic and kinetic energy of the transverse perturbations to the background magnetic field in a subdomain where $x,y \in \left[ -2.03 R, 2.03 R  \right]$. Then, we average the energy in $y$- and $z$-directions and  retain the dependence in the $x$-direction only. We denote this averaged energy as $\overline{E}(x,t)$. After that, we apply the continuous Fourier transform, which is discretized due to the limited numerical resolution as
\begin{align}
\overline{E}_F(k_\perp,t)\approx\frac{\Delta x}{\sqrt{2\pi}}\exp{\left(-i k_\perp x_{0}\right)} \sum^{N-1}_{m=0} \overline{E}(x,t)\exp{\left(-\frac{2\pi imk_\perp}{N}\right)},
\label{fftdiscr}
\end{align}
where $N$ is the number of samples, $k_\perp$ is the perpendicular wave number, and $\Delta x $ and  $x_{0}$ are the spatial resolution and the upper limit of the selected subdomain in the $x$-direction, respectively. The summatory in Eq.~(\ref{fftdiscr}) is exactly the 1D discrete Fourier transform, which we computed using the FFT algorithm \citep{Cooley65}. Finally, we calculate the modulus of $\overline{E}_F(k_\perp,t)$,  and normalize it with respect to the maximum value in the spectrum at each time. Moreover, to compare with linear results, we repeat this computation, but with the analytic formulas in the linear regime. For simplicity, this analysis was done in the thin-layer case alone. The results of the Fourier transform are plotted in Fig.~\ref{ffttransporthin} corresponding to the final frame of the simulation. In Fig.~\ref{ffttransporthin}, with a vertical line, we show the maximum theoretical wave number across the loop, $k_{\rm max}$, generated by phase mixing (we used Eq.~(\ref{effwnu})). The results of the analytic spectrum for $k_\perp >k_{\rm max}$ are unphysical and must be ignored because phase mixing cannot generate such large wave numbers \citep{Mann95}.  This is just a numerical artifact caused by the FFT routine when it is forced to be extended to larger wave numbers in order to obtain the same wave-number range as in the numerical spectrum.

As expected, we verified that at $t=0$ and during the initial simulation frames, analytical and numerical spectrums coincide (not shown here). This is the case until the onset of the KHi. Then, the analytical and numerical spectra start to differ for sufficiently high wave numbers. Interestingly, they differ for a wave number that is somewhat smaller than  $k_{\rm max}$. This clearly confirms that the development of turbulence produces much smaller scales (higher waver numbers) than what phase mixing alone is capable of. For $k_\perp>k_{\rm max}$, the amplitude of the numerical spectrum is much larger than that of the analytic spectrum. In fact, as mentioned before, the amplitude of the analytic spectrum should tend to zero for $k_\perp>k_{\rm max}$. Therefore, in the full nonlinear results there is a faster cascading of energy to small scales than in the linear regime.

The numerical simulations show that the KHi drives the loop to a turbulent state and that turbulence speeds up the  energy transport from large to small scales. One question that may arise is whether the energy cascading obtained here agrees with what is expected under conditions of turbulence. According to turbulence theory \citep[see, e.g.,][]{Chorin94,Frisch95,Pope2000,Schnack09}, spectra can be divided into two main regions: the energy-containing range and the universal equilibrium range, which, in turn, contains the inertial subrange and the dissipation range. The scaling of the energy with the wave number in the inertial subrange can give us the information we are looking for.

The  wave number that separates the inertial subrange from the energy-containing range is difficult to estimate because it depends on the characteristics of the problem  \citep{Pope2000,Bluteau11}. A typical spectrum for a driven system should contain a region of small wave numbers where energy increases with the wave number before it decreases in the inertial subrange \citep{Biskmap03}. However, we do not have this increase in our spectrum because we did not use a driver, which would input energy continuously. Instead, we imposed an initial perturbation.  To make sure that the considered wave numbers fell entirely within the inertial subrange, we considered $k_\perp >k_{\rm max}$. This choice also excludes the effect of linear phase mixing. On the other hand, the dissipation range, where the energy is dissipated, cannot be studied in ideal MHDs. Under coronal conditions, the critical  wave number between the inertial subrange and the dissipation range should be larger than the wave numbers considered in Fig.~\ref{ffttransporthin}. Nevertheless, although small, some numerical diffusion is at work. To eliminate any effect numerical dissipation may have, we excluded wave numbers higher than an estimated numerical diffusion wave number, $k_{\rm diff}$.  Therefore, we assume that the inertial subrange approximately corresponds to $k_{\rm max}  < k_\perp  < k_{\rm diff}$.  For the case of Fig.~\ref{ffttransporthin}, we have $k_{\rm max} R \approx 105,$ and we visually estimated $k_{\rm diff} R \approx 500$ from an evident slope change in the spectrum at high wave numbers (see below).

In order to determine the approximate slope of the numerical spectrum in the inertial subrange, we performed a least-squares linear fit in log-log scale. The result of the best fit is overplotted in Fig.~\ref{ffttransporthin}. The adjusted slope is $-2.07 \pm 0.02$. On the other hand, an equivalent fit performed in  Fig.~\ref{ffttransporthin} for  $k_\perp  > k_{\rm diff}$ reveals a steeper slope of $-3.28 \pm 0.07$, which is consistent with the assumption that for  $k_\perp  > k_{\rm diff}$ numerical diffusion plays a role in this high-wave-number range.

We note that the fit slope in the inertial subrange may change if the estimated boundary wave numbers, namely $k_{\rm max}$ and $k_{\rm diff}$, are modified. We are aware that the choice of these boundary wave numbers is not free from a certain degree of arbitrariness. In spite of this and considering the limitations of this rough estimation,  the obtained slope of $\approx -2$ differs from the so-called Kolmogorov law of $-5/3$ \citep[see e.g.,][]{Kolmogorov41,Frisch95,Pope2000}. The Kolmogorov scaling law is a universal property in hydrodynamic turbulence  and also appears in strong MHD turbulence \citep[see][]{Biskmap03,Schnack09}. The obtained slope  is also different for the Iroshnikov-Kraichnan law of $-3/2$ because of isotropic Alfv\'enic turbulence \citep{Iroshnikov1964,Kraichnan1965}. We note that neither the condition of isotropy  nor that of strong turbulence are  satisfied in our simulations, which may explain why we obtain a different slope. As discussed by \citealt{Nazarenko2011} and \citealt{Schekochihin2012}; see also, \citealt{Ng1997,galtier2000}, in the case of  anisotropic  weak Alfv\'enic turbulence that evolves nearly 2D in the plane transverse to the background magnetic field   with $k_\perp \gg k_\parallel$, as it is indeed the case of our simulations, the energy spectrum should scale as $\sim k_\perp^{-2}$. This scaling law is compatible with the approximate slope we fit in  Fig.~\ref{ffttransporthin} for $k_{\rm max}  < k_\perp  < k_{\rm diff}$. However, we must warn the reader that the -2 power law discussed in \citet{Nazarenko2011} and \citet{Schekochihin2012} is obtained for unbalanced turbulence arising from wave-wave interaction, while the turbulence in our case is driven by the KHi. Therefore, caution is needed with the comparison of the power law (A. Hillier, private communication).

\subsection{Effective Reynolds number}

A consequence of the faster generation of small scales owing to turbulence is that the dissipative scales should be reached much before than the theory of linear phase mixing predicts \citep[see][]{terradasarregui2018}. A way to further quantify the
differences between linear and nonlinear results is to estimate the effective Reynolds number in the simulations as \begin{equation}
R_{\rm eff}=\frac{\left|\rho\left[\frac{\partial \mathbf{v}}{\partial t}+\left(\mathbf{v} \cdot \nabla\right)\mathbf{v}\right]\right|}{\left|\rho\nu\left[\nabla^{2}\mathbf{v}+\frac{1}{3}\nabla\left(\nabla \cdot \mathbf{v}\right)\right]\right|}.
\label{effrey}
\end{equation}
The numerator of Eq.~(\ref{effrey}) is the magnitude of the inertial term, whereas the denominator is the magnitude of the viscous force, with $\nu$ the kinematic viscosity \citep[see, e.g.,][]{Priest12}. Using $R$ and $v_{\rm A,e}$ as characteristic length and velocity scales, Eq.~(\ref{effrey}) gives
\begin{equation}
R_{\rm eff}=\frac{v_{A,e} R}{\nu} \approx 10^{12}
\label{nuu}
\end{equation}
for typical values of the parameters in the corona. We recall that we performed ideal MHD simulations, so the viscous term is absent from the equations we actually solved. Assuming that the effect of viscosity, if included, would regardless be small during the considered simulation times because we are still far from the dissipative scales, we can approximately study how the Reynolds number would evolve in our simulations.

\begin{figure}[htbp]
\centering
\resizebox{\hsize}{!}{\includegraphics{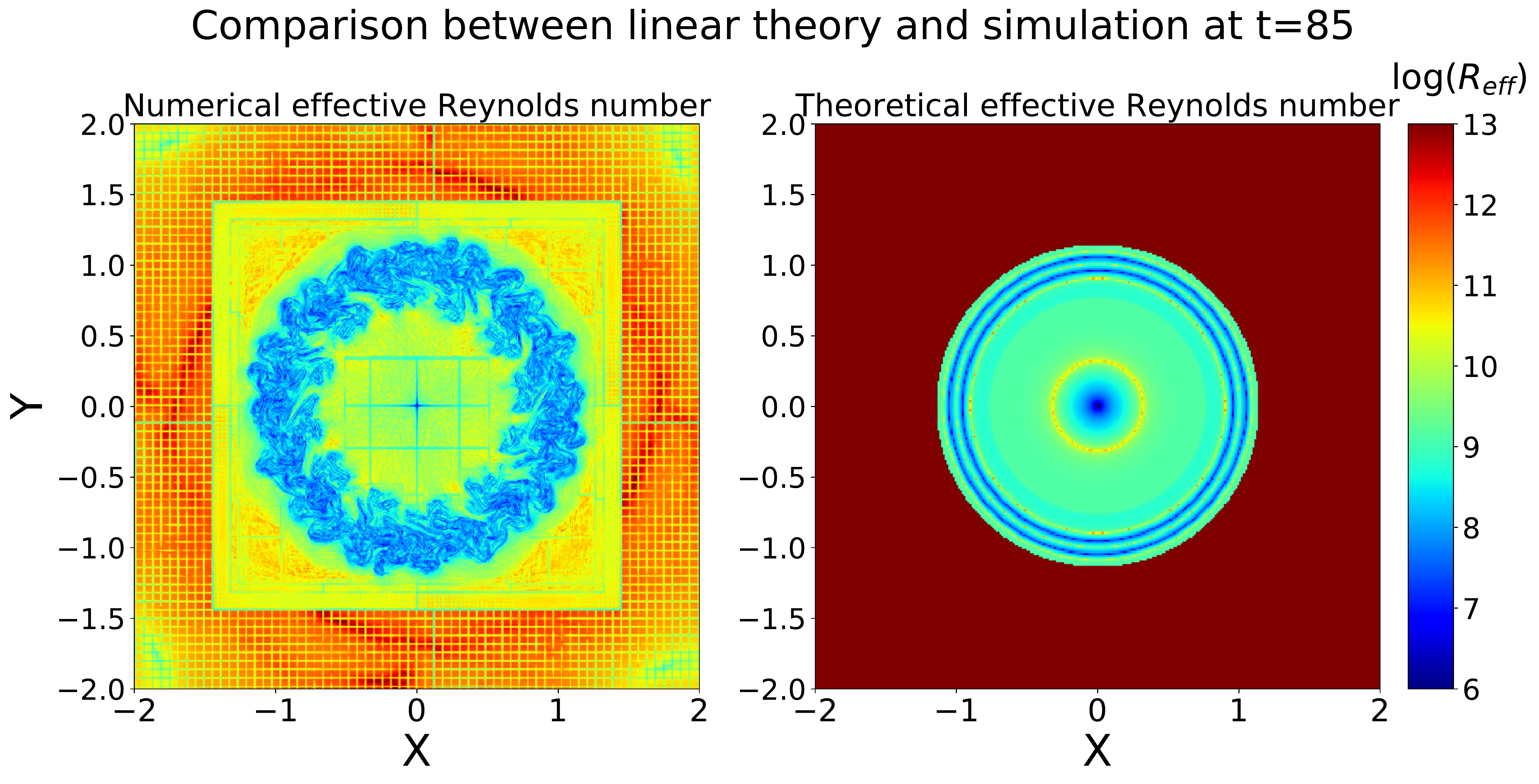}}
\resizebox{\hsize}{!}{\includegraphics{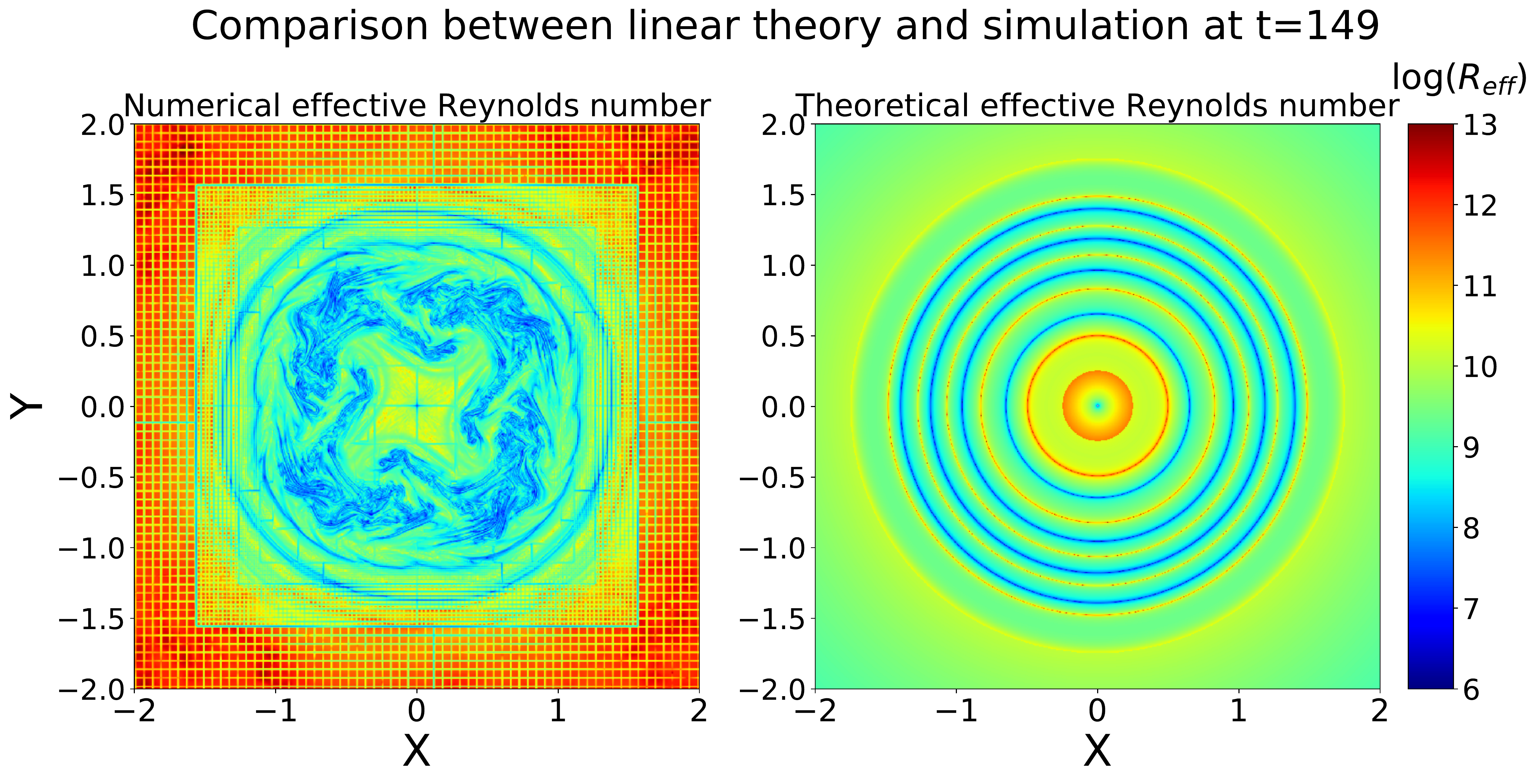}}
\caption{{\em Top:} Numerical (left) and analytic (right) effective Reynolds number in logarithmic scale at the end time of the simulation in the thin-layer case. These results are calculated in a cross-sectional cut at $z=0$. The straight lines seen in the numerical result are visualization artifacts due to the AMR scheme. These artifacts are not present in the actual simulation data. {\em Bottom:} Same as top panels, but for the thick-layer case.}
\label{logreynoldsthin}
\end{figure}

Figure~\ref{logreynoldsthin} shows a cross-sectional cut of the effective Reynolds number at the tube center at the end of the simulation times for both thin-layer and thick-layer cases. In order to compare with the linear theory, we also include the results that we obtained using the theoretical expressions of the velocity in the quasi-linear analysis. We also observed the evolution of $R_{\rm eff}$ with the simulation time (not shown here) before the time plotted in Fig.~\ref{logreynoldsthin}.  As expected, numerical and theoretical effective Reynolds numbers are initially similar. As time increases, phase mixing is developed in the nonuniform layer, and $R_{\rm eff}$ varies in the form of concentric rings between  minimums and  maximums. These minimums of $R_{\rm eff}$ are very local, that is, they only occur in very thin regions within the nonuniform layer.

When the simulations reach the nonlinear regime, the effective numerical Reynolds number dramatically decreases in the nonuniform layer due to the KHi. The formation of eddies associated with the KHi greatly increases the denominator of Eq.~(\ref{effrey}), that is, the size of the hypothetical viscous term. As turbulence develops,  low values of $R_{\rm eff}$ are found within the entire nonuniform layer. The decrease of $R_{\rm eff}$ is no longer a very local phenomenon, as it happens when phase mixing is operating alone. At the final time of the simulations, the values of $R_{\rm eff}$ are  as low as $R_{\rm eff} \sim 10^6$, that is, six orders of magnitude lower than at $t=0$. The thick-layer case, where the KHi is less developed than in the thin-layer case, shows an intermediate step in which the dramatic decrease of $R_{\rm eff}$ due to turbulence only occurs in the inner part of the nonuniform region, while in the outermost part of the layer,  phase mixing is still at work and larger values of $R_{\rm eff}$ are found.

\subsection{Parameter survey}
\label{parstudy}
 
To this point, we considered fixed values for the initial amplitude, $\varepsilon$, the width of the transition region, $l/R$, the density contrast between the core of the flux tube and the external medium, $ \rho_{\rm i}/\rho_{\rm e} $, and the loop length, $ L/R $. Here, we explore the effect that each one of these four parameters has on the excitation of the KHi during the nonlinear evolution of the torsional oscillations. To do so, we ran different simulations by keeping three of the parameters fixed and varying the remaining one. We performed 15 different simulations. For each simulation, we computed the vorticity squared integrated over the whole computational domain using Eq.~(\ref{vortsq}) and estimated the value of $\tau_{\rm KH}$ in each case.

\begin{figure}[!tbp]
\centering
\resizebox{\hsize}{!}{\includegraphics{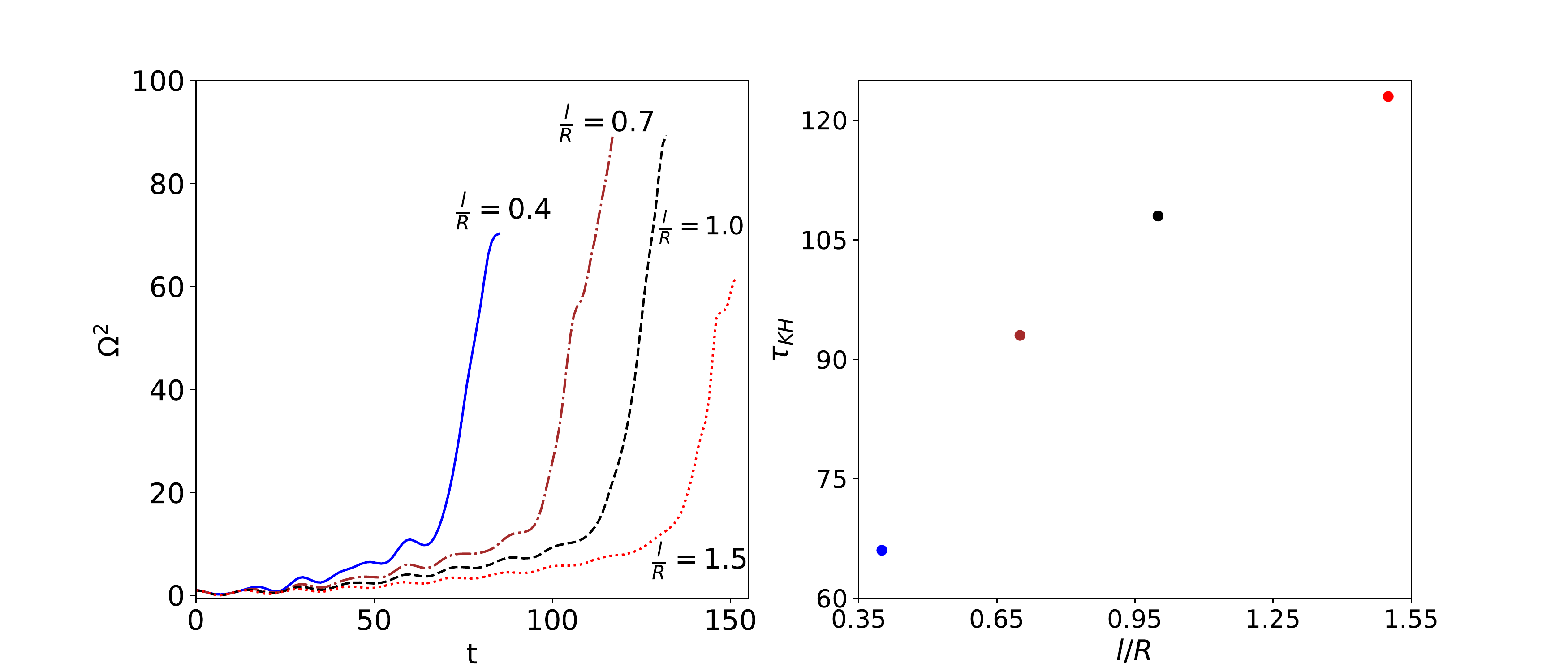}}
\caption{\textit{Left panel}: Vorticity squared integrated over the whole computational domain as a function of the simulation time for $\varepsilon=0.1$, $ \rho_{i}/\rho_{e}=2.0 $, $L/R=10$, and $l/R=\left\lbrace 0.4,0.7,1.0,1.5 \right\rbrace $. The curves are normalized to the integrated vorticity squared at $t=0$. \textit{Right panel}: Estimated $\tau_{\rm KH}$ as a function of the width of the transition region, $l/R$.}
\label{studywidth}
\end{figure}

\begin{figure}[!tbp]
\centering
\resizebox{\hsize}{!}{\includegraphics{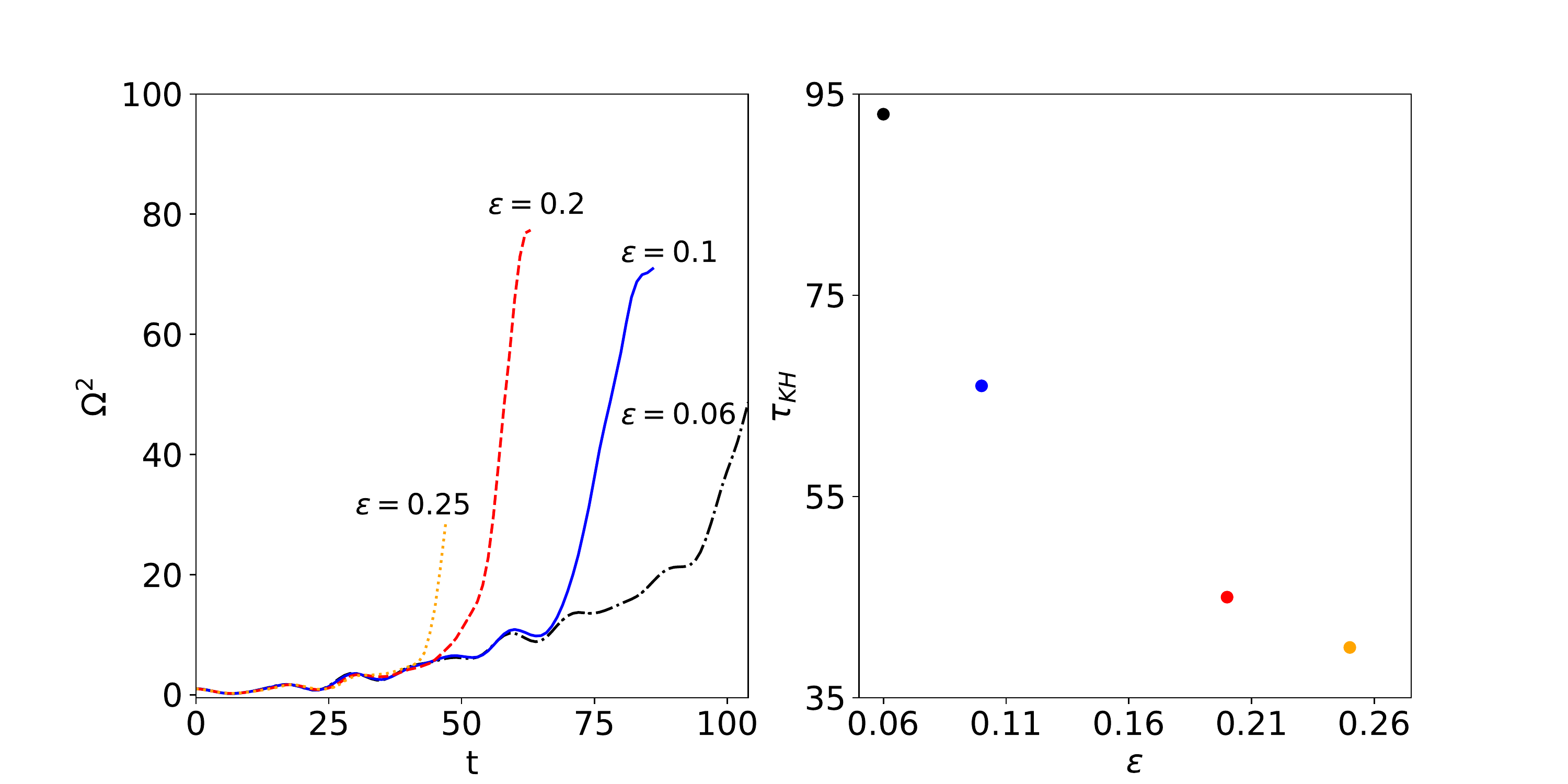}}
\caption{\textit{Left panel}: Same as left panel of Fig.~\ref{studywidth}, but now the fixed parameters are $ l/R=0.4 $, $L/R=10$, and $ \rho_{i}/\rho_{e}=2.0 $, and we vary $\varepsilon=\left\lbrace 0.06, 0.1, 0.2, 0.25 \right\rbrace $. \textit{Right panel}: Estimated $\tau_{\rm KH}$ as a function of the amplitude of the velocity  perturbation.}
\label{studyamplitude}
\end{figure}

\begin{figure}[!tbp]
\centering
\resizebox{\hsize}{!}{\includegraphics{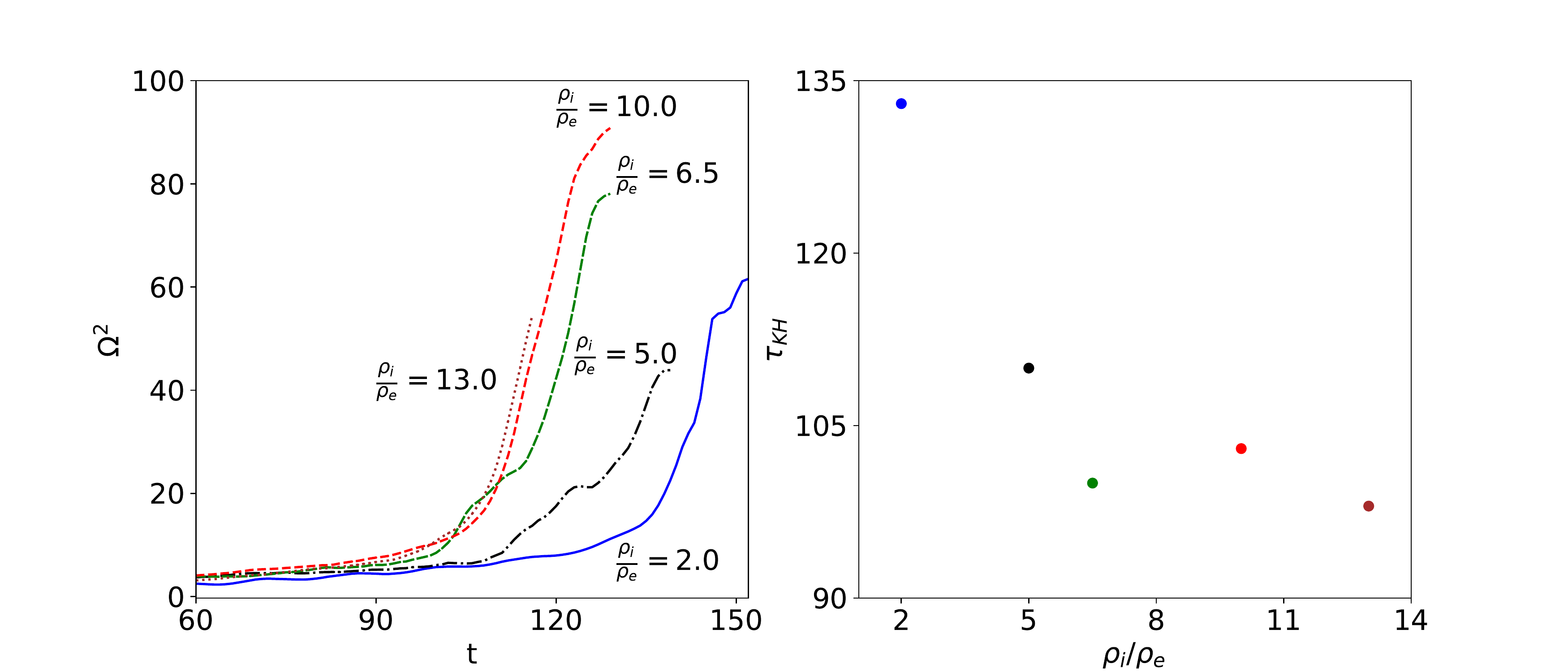}}
\caption{\textit{Left panel}: Same as left panel of Fig.~\ref{studywidth}, but now the fixed parameters are $ \varepsilon=0.1 $, $L/R=10$, and $ \rho_{i}/\rho_{e}=2.0 $, and we vary $\rho_{\rm i}/\rho_{\rm e}=\left\lbrace 2, 5, 6.5, 10, 13 \right\rbrace $. \textit{Right panel}: Estimated $\tau_{\rm KH}$ as a function of the density contrast between the core of the flux tube and the external medium.}
\label{studydensity}
\end{figure}

\begin{figure}[!tbp]
\centering
\resizebox{\hsize}{!}{\includegraphics{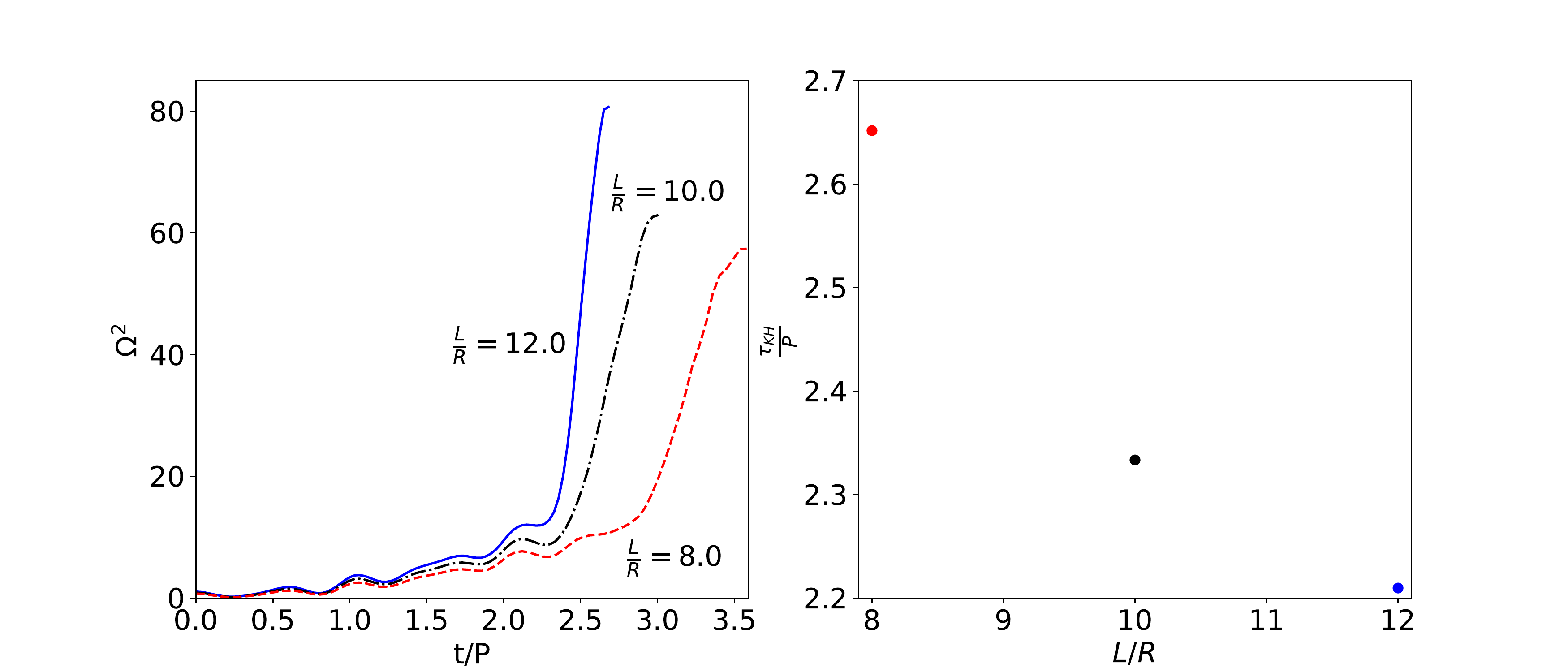}}
\caption{\textit{Left panel}:  Same as left panel of Fig.~\ref{studywidth}, but now the fixed parameters are $ \varepsilon=0.1 $, $l/R=0.4 $, and $ \rho_{i}/\rho_{e}=2.0 $, and we vary $L/R=\left\lbrace 8.0,10.0,12.0 \right\rbrace $. \textit{Right panel}: Estimated $\tau_{\rm KH}$  as a function of the loop length. We note that in these panels time is normalized to the period of the internal torsional Alfv\'{e}n wave, $ P = 2L / v_{\rm A,i} $.}
\label{studylength}
\end{figure}

Figure~\ref{studywidth} shows the results of the parameter study for fixed $\varepsilon=0.1$ , $L/R=10$, and $ \rho_{i}/\rho_{e}=2 $, and varying $l/R=\left\lbrace 0.4,0.7,1.0,1.5 \right\rbrace $. The left panel shows the evolution of integrated vorticity squared as a function of time, where we can clearly identify the linear phase and the subsequent  change of slope at $t=\tau_{\rm KH}$ due to the {\em \emph{global onset}} of the KHi. The right panel shows the estimated values of $\tau_{\rm KH}$ for every simulation with different $l/R$. As expected, we find that the onset of KHi is delayed as $l/R$ increases, and, in the linear regime, the slope of the integrated vorticity squared depends on the value of $l/R$.  This is consistent with the fact that the phase-mixing pace depends on the inhomogeneity length scale: it operates faster (slower) when $l/R$ decreases (increases). The strong enough shear flows needed to trigger the KHi occur at a later time as $l/R$ grows. Conversely, after the KHi is triggered, the slope of the integrated vorticity squared is approximately the same for all values of $l/R$, which points out that the nonlinear evolution of the KHi is less affected by the inhomogeneity length scale.

Figure~\ref{studyamplitude} shows the same analysis, but when $l/R=0.4$, $L/R=10$, and $ \rho_{i}/\rho_{e}=2 $ are kept fixed, and we vary $\varepsilon=\left\lbrace 0.06, 0.1, 0.2, 0.25 \right\rbrace $. As expected, we find that the larger the perturbation amplitude, the earlier the KHi sets in. This is so because larger  shear flows are present and the nonlinear regime is more rapidly reached as the amplitude increases. We notice that, as should be the case, all curves in the left panel of Fig.~\ref{studyamplitude} superimpose in the linear regime before the onset of the KHi. 

Finally, Fig.~\ref{studydensity} displays the results when $l/R=1.5$, $L/R=10$, and $ \varepsilon=0.1 $ are kept fixed and we vary $\rho_{\rm i}/\rho_{\rm e}=\left\lbrace 2, 5, 6.5, 10, 13 \right\rbrace $. The dependence of $\tau_{\rm KH}$ on the density contrast is not as clear as for the other two parameters. For low-density contrasts, $\tau_{\rm KH}$ decreases when $\rho_{\rm i}/\rho_{\rm e}$ increases. However, $\tau_{\rm KH}$ appears to saturate to a roughly constant value for higher density contrasts. The initial decrease of $\tau_{\rm KH}$ with $\rho_{\rm i}/\rho_{\rm e}$ is consistent with the fact that a larger density contrast implies a more abrupt Alfv\'{e}n speed variation within the nonuniform layer and so a faster development of phase mixing. This seems to be true up to a certain value of the contrast, but the precise value of $\rho_{\rm i}/\rho_{\rm e}$ becomes irrelevant once it is large enough. We speculate that this behavior may be caused by the functional dependence of the Alfv\'en speed with the density, namely $v_{\rm A} \sim \rho^{-1/2}$.

 The dependence of $\tau_{\rm KH}$ with $l/R$ and $\varepsilon$ qualitatively agrees with the behavior predicted by the approximate critical time of \citet{Browning} in the strong phase-mixing limit (see Eq.~(\ref{tcri})). However, Eq.~(\ref{tcri}) is independent of the density contrast. The result that $\tau_{\rm KH}$ turns independent of $\rho_{\rm i}/\rho_{\rm e}$ for large values of this parameter may indicate that  the strong phase-mixing limit of \citet{Browning} becomes more applicable as the density contrast increases.

We recall that for $\rho_{\rm i}/\rho_{\rm e} = 2,$ the periods of the internal and external Alfv\'en waves are $20\sqrt{2} \approx 28.3$ and 20 time units, respectively, when $L/R = 10$. The values of $\tau_{\rm KH}$ obtained in the above parameter study correspond to few oscillation periods. This means that the KHi is triggered in a relatively short timescale once the torsional oscillation begins. However, the period depends on the loop length and may also have some influence on the KHi onset time. \citet{Terradas08} and \citet{Antolin14} showed that the loop length influences the onset of the KHi driven by the kink mode, so it is worth exploring the role of this parameter in the case of torsional waves. Figure~\ref{studylength} displays the integrated vorticity analysis for fixed $\varepsilon=0.1$ , $l/R=0.4$, and $ \rho_{i}/\rho_{e}=2 $, and varying $L/R=\left\lbrace 8.0,10.0,12.0 \right\rbrace $. Since now the period of the torsional Alfv\'{e}n waves are different for each simulation,  to fairly compare these simulations, time is normalized to the period of the internal torsional Alfv\'{e}n wave, namely $ P = 2L / v_{\rm A,i} $. By doing so, we eliminate the implicit dependence of the simulation time with the loop length. Indeed, the results show that $\tau_{\rm KH}/P$ is weakly affected by the loop length. We find that the longer the loop, the earlier the relative onset time, but the differences are modest for the considered values of $L/R$. In all cases, the KHi is triggered for times between two and three internal periods for the considered parameters. We must recall that the loops considered here are shorter than those typically observed in the corona. If the trend shown in Fig.~\ref{studylength} holds for longer loops, the important conclusion would be that the KHi can develop in a realistically long loop in a timescale equal to a few periods of the torsional wave.

\section{Conclusions}

We studied the nonlinear evolution of phase-mixed torsional Alfv\'en waves in a low-$\beta$ coronal flux tube model with a constant axial magnetic field, which is line tied at the photosphere. This model consists of a tube with a uniform  core, a transition layer where the density decreases continuously, and an external medium with uniform density.

The longitudinally fundamental  mode of standing torsional Alfv\'{e}n waves is excited at $t=0$. First, the temporal evolution of the waves undergoes a quasi-linear phase that is well predicted by approximate analytical theory. Phase mixing evolves within the nonuniform layer by developing spatial oscillations of the azimuthal components of velocity and magnetic field that lead to the generation of smaller scales \citep[see, e.g.,][]{HeyvaertPriest83}. As a result, vorticity and current density increase in the system. In addition, the nonlinear coupling of Alfv\'{e}n waves with slow magnetoacoustic waves through the ponderomotive force \citep[see, e.g.,][]{Hollweg71} causes longitudinal flows. 

Later in the evolution, the azimuthal shear flows generated by phase mixing trigger the KHi \citep{Browning}.  From that time onwards, the numerical results deviate significantly from those of quasi-linear analytic theory. Once the KHi is triggered, higher azimuthal wave numbers than the torsional mode appear in the system. As time increases, the KHi dramatically increases the total values of vorticity and current density and greatly enhances the development of small scales first initiated by phase mixing during the linear phase. The onset time of the KHi can be estimated from the evolution of the vorticity squared integrated over the whole computational box. Although this onset time depends on various model parameters, the KHi is typically triggered after few oscillation periods of the torsional waves. In particular, the parameter study showed that the wider the transition region, the later the KHi starts. Thus, a wide enough transition region can delay the onset and growth of KHi (see \citealt{Terradas18} for a similar result in the case of kink modes). In turn, the larger the velocity amplitude, the earlier the KHi is triggered because the system reaches the nonlinear regime faster. Regarding the density contrast, we find that for a small contrast, the higher the density contrast, the earlier the onset of the KHi. For sufficiently large contrasts, the onset time becomes roughly independent of this parameter. Finally, it is found that the longer the loop, the earlier the KHi is triggered relatively to the oscillation period of the torsional waves.

The KHi onset time estimated from the vorticity evolution always overestimates the critical times predicted analytically by  \citet{Browning}. The reason for this discrepancy is that the critical time of \citet{Browning} corresponds to the time at which the KHi is locally triggered within the nonuniform layer, while the onset time obtained from the vorticity evolution should be understood as a time for which the KHi has evolved enough to globally impact on the whole flux tube dynamics.

The development of the KHi drives the flux tube to a turbulent state. We find that the energy associated with the perturbations cascades with the perpendicular wavenumber following a -2 power law. Spatial scales much smaller than those generated by phase mixing alone are present during the turbulent phase. As a consequence, the effective Reynolds number decreases in the system much faster than what linear phase mixing predicts, meaning that the dissipative scales are  approached much quicker \citep{terradasarregui2018, Ebrahimi20}. Our ideal MHD simulations stop before such dissipative scales, which are reached in the corona. Those dissipative scales would eventually be reached at a later time, should turbulence continue to naturally evolve in the flux tube.

 After turbulence has developed, the energy cascade to small scales speeds up considerably. Although the considered four-level AMR scheme is able to deal with those small scales during the initial stages of turbulence, it eventually fails to capture the fast generation of fine structures. We stopped our simulations precisely at that time to guarantee the physical validity of the results. For later times, numerical dissipation takes a predominant role and artificially affects the evolution. This highlights how crucial using sufficiently high spatial resolutions is to correctly resolve the fine dynamics associated with the nonlinear, turbulent evolution of the waves. 

In connection with the recent results of \citet{Soler20}, the turbulent evolution of the torsional oscillations may provide a way to dissipate the large  energy budget transmitted from the photosphere.  \citet{Soler20} conclude that wave energy dissipation is very inefficient in the linear regime. However, the turbulence discussed here may significantly enhance the heating rates \citep[see, e.g.,][]{vanBallegooijen2011,Asgari2012,vanBallegooijen2017,Hillier2020}. The results obtained here in this simple flux tube model open the door to more elaborate future studies. In the solar corona, where flux tubes are naturally inhomogeneous, this type of wave-mediated turbulence should occur, and there is much room to explore \citep[see also][]{Magyar2019}.  Additional effects to investigate could be, for instance, nonideal terms to study the actual energy dissipation, adding magnetic twist, and the evolution of torsional waves in curved coronal loops.

\begin{acknowledgement}
We acknowledge the support from grant AYA2017-85465-P (MINECO/AEI/FEDER, UE). SD acknowledges the support from the Spanish Ministry of Science and Innovation for the predoctoral FPI felloship  PRE2018-084223. We thank Jaume Terradas for his technical assistance. We also thank the Universitat de les Illes Balears for the use of the Foner cluster. For the simulation data analysis we have used VisIT \citep{visit} and Python 3.5. In particular, we have used Matplotlib \citep{Huntermatplotlib}, Scipy \citep{scipycitation}, and Numpy \citep{numpycitation}. We are thankful to B. Vaidya and his contributors for the tool pyPLUTO. Finally, we thank Andrew Hillier, Tom Van Doorsselaere, and the anonymous referee for helpful comments.
\end{acknowledgement}

\bibpunct{(}{)}{;}{a}{}{,} 
\bibliographystyle{aa} 
\bibliography{thebib} 

\begin{thebibliography}{111}
\expandafter\ifx\csname natexlab\endcsname\relax\def\natexlab#1{#1}\fi

\bibitem[{{Alfv{\'e}n}(1942)}]{Alf42}
{Alfv{\'e}n}, H. 1942, \nat, 150, 405

\bibitem[{{Antolin} {et~al.}(2015){Antolin}, {Okamoto}, {De Pontieu},
  {Uitenbroek}, {Van Doorsselaere}, \& {Yokoyama}}]{Antolin15}
{Antolin}, P., {Okamoto}, T.~J., {De Pontieu}, B., {et~al.} 2015, \apj, 809, 72

\bibitem[{{Antolin} \& {Van Doorsselaere}(2019)}]{Antolin19}
{Antolin}, P. \& {Van Doorsselaere}, T. 2019, Frontiers in Physics, 7, 85

\bibitem[{{Antolin} {et~al.}(2014){Antolin}, {Yokoyama}, \& {Van
  Doorsselaere}}]{Antolin14}
{Antolin}, P., {Yokoyama}, T., \& {Van Doorsselaere}, T. 2014, \apjl, 787, L22

\bibitem[{{Arregui} {et~al.}(2011){Arregui}, {Soler}, {Ballester}, \&
  {Wright}}]{arregui2011}
{Arregui}, I., {Soler}, R., {Ballester}, J.~L., \& {Wright}, A.~N. 2011, \aap,
  533, A60

\bibitem[{{Aschwanden} \& {Wang}(2020)}]{Aschwanden20}
{Aschwanden}, M.~J. \& {Wang}, T. 2020, \apj, 891, 99

\bibitem[{{Asgari-Targhi} \& {van Ballegooijen}(2012)}]{Asgari2012}
{Asgari-Targhi}, M. \& {van Ballegooijen}, A.~A. 2012, \apj, 746, 81

\bibitem[{{Ballester} {et~al.}(2020){Ballester}, {Soler}, {Terradas}, \&
  {Carbonell}}]{Pep20}
{Ballester}, J., {Soler}, R., {Terradas}, J., \& {Carbonell}, M. 2020, \aap,
  641, 17

\bibitem[{{Barbulescu} {et~al.}(2019){Barbulescu}, {Ruderman}, {Van
  Doorsselaere}, \& {Erd{\'e}lyi}}]{Barbulescu19}
{Barbulescu}, M., {Ruderman}, M.~S., {Van Doorsselaere}, T., \& {Erd{\'e}lyi},
  R. 2019, \apj, 870, 108

\bibitem[{{Berger} {et~al.}(2010){Berger}, {Slater}, {Hurlburt}, {Shine},
  {Tarbell}, {Title}, {Lites}, {Okamoto}, {Ichimoto}, {Katsukawa}, {Magara},
  {Suematsu}, \& {Shimizu}}]{Berger10}
{Berger}, T.~E., {Slater}, G., {Hurlburt}, N., {et~al.} 2010, \apj, 716, 1288

\bibitem[{Biskamp(2003)}]{Biskmap03}
Biskamp, D. 2003, Magnetohydrodynamic Turbulence (Cambridge University Press)

\bibitem[{Bluteau {et~al.}(2011)Bluteau, Jones, \& Ivey}]{Bluteau11}
Bluteau, C.~E., Jones, N.~L., \& Ivey, G.~N. 2011, Limnology and Oceanography:
  Methods, 9, 302

\bibitem[{{Browning} \& {Priest}(1984)}]{Browning}
{Browning}, P.~K. \& {Priest}, E.~R. 1984, \aap, 131, 283

\bibitem[{{Cargill} \& {de Moortel}(2011)}]{Cargill11}
{Cargill}, P. \& {de Moortel}, I. 2011, \nat, 475, 463

\bibitem[{Chandrasekhar(1961)}]{Chandra}
Chandrasekhar, S. 1961, Hydrodynamic and Hydromagnetic Stability, International
  series of monographs on physics (Clarendon Press)

\bibitem[{{Charbonneau} \& {MacGregor}(1995)}]{Charbonneau95}
{Charbonneau}, P. \& {MacGregor}, K.~B. 1995, \apj, 454, 901

\bibitem[{Childs {et~al.}(2012)Childs, Brugger, Whitlock, Meredith, Ahern,
  Pugmire, Biagas, Miller, Harrison, Weber, Krishnan, Fogal, Sanderson, Garth,
  Bethel, Camp, Rubel, Durant, Favre, \& Navratil}]{visit}
Childs, H., Brugger, E., Whitlock, B., {et~al.} 2012, High Performance
  Visualization-Enabling Extreme-Scale Scientific Insight, 357

\bibitem[{Chorin(1994)}]{Chorin94}
Chorin, A. 1994, Vorticity and Turbulence, Applied Mathematical Sciences
  (Springer)

\bibitem[{Cooley \& Tukey(1965)}]{Cooley65}
Cooley, J.~W. \& Tukey, J.~W. 1965, Mathematics of computation, 19, 297

\bibitem[{{Cranmer}(2009)}]{Cranmer09}
{Cranmer}, S.~R. 2009, Living Reviews in Solar Physics, 6, 3

\bibitem[{{Cranmer} \& {van Ballegooijen}(2005)}]{Crammer05}
{Cranmer}, S.~R. \& {van Ballegooijen}, A.~A. 2005, \apjs, 156, 265

\bibitem[{{De Moortel} {et~al.}(2000){De Moortel}, {Hood}, \&
  {Arber}}]{Moortel00}
{De Moortel}, I., {Hood}, A.~W., \& {Arber}, T.~D. 2000, \aap, 354, 334

\bibitem[{{De Moortel} {et~al.}(2014){De Moortel}, {McIntosh}, {Threlfall},
  {Bethge}, \& {Liu}}]{DeMoortel2014}
{De Moortel}, I., {McIntosh}, S.~W., {Threlfall}, J., {Bethge}, C., \& {Liu},
  J. 2014, \apjl, 782, L34

\bibitem[{{De Pontieu} {et~al.}(2012){De Pontieu}, {Carlsson}, {Rouppe van der
  Voort}, {Rutten}, {Hansteen}, \& {Watanabe}}]{depontieu2012}
{De Pontieu}, B., {Carlsson}, M., {Rouppe van der Voort}, L.~H.~M., {et~al.}
  2012, \apjl, 752, L12

\bibitem[{{De Pontieu} {et~al.}(2007){De Pontieu}, {McIntosh}, {Carlsson},
  {Hansteen}, {Tarbell}, {Schrijver}, {Title}, {Shine}, {Tsuneta}, {Katsukawa},
  {Ichimoto}, {Suematsu}, {Shimizu}, \& {Nagata}}]{Depon07}
{De Pontieu}, B., {McIntosh}, S.~W., {Carlsson}, M., {et~al.} 2007, Science,
  318, 1574

\bibitem[{{De Pontieu} {et~al.}(2014){De Pontieu}, {Rouppe van der Voort},
  {McIntosh}, {Pereira}, {Carlsson}, {Hansteen}, {Skogsrud}, {Lemen}, {Title},
  {Boerner}, {Hurlburt}, {Tarbell}, {Wuelser}, {De Luca}, {Golub}, {McKillop},
  {Reeves}, {Saar}, {Testa}, {Tian}, {Kankelborg}, {Jaeggli}, {Kleint}, \&
  {Martinez-Sykora}}]{Depon14}
{De Pontieu}, B., {Rouppe van der Voort}, L., {McIntosh}, S.~W., {et~al.} 2014,
  Science, 346, 1255732

\bibitem[{{Dedner} {et~al.}(2002){Dedner}, {Kemm}, {Kr{\"o}ner}, {Munz},
  {Schnitzer}, \& {Wesenberg}}]{Dedner02}
{Dedner}, A., {Kemm}, F., {Kr{\"o}ner}, D., {et~al.} 2002, Journal of
  Computational Physics, 175, 645

\bibitem[{{Ebrahimi} {et~al.}(2020){Ebrahimi}, {Soler}, \&
  {Karami}}]{Ebrahimi20}
{Ebrahimi}, Z., {Soler}, R., \& {Karami}, K. 2020, \apj, 893, 157

\bibitem[{{Edwin} \& {Roberts}(1983)}]{Edwin83}
{Edwin}, P.~M. \& {Roberts}, B. 1983, \solphys, 88, 179

\bibitem[{{Fedun} {et~al.}(2011){Fedun}, {Shelyag}, {Verth}, {Mathioudakis}, \&
  {Erd{\'e}lyi}}]{fedun2011}
{Fedun}, V., {Shelyag}, S., {Verth}, G., {Mathioudakis}, M., \& {Erd{\'e}lyi},
  R. 2011, Annales Geophysicae, 29, 1029

\bibitem[{{Foullon} {et~al.}(2011){Foullon}, {Verwichte}, {Nakariakov},
  {Nykyri}, \& {Farrugia}}]{Foullon11}
{Foullon}, C., {Verwichte}, E., {Nakariakov}, V.~M., {Nykyri}, K., \&
  {Farrugia}, C.~J. 2011, \apjl, 729, L8

\bibitem[{Frisch(1995)}]{Frisch95}
Frisch, U. 1995, Turbulence: The Legacy of A. N. Kolmogorov (Cambridge
  University Press)

\bibitem[{{Galtier} {et~al.}(2000){Galtier}, {Nazarenko}, {Newell}, \&
  {Pouquet}}]{galtier2000}
{Galtier}, S., {Nazarenko}, S.~V., {Newell}, A.~C., \& {Pouquet}, A. 2000,
  Journal of Plasma Physics, 63, 447

\bibitem[{{Goossens} {et~al.}(2012){Goossens}, {Andries}, {Soler}, {Van
  Doorsselaere}, {Arregui}, \& {Terradas}}]{Goossens12}
{Goossens}, M., {Andries}, J., {Soler}, R., {et~al.} 2012, \apj, 753, 111

\bibitem[{{Goossens} {et~al.}(2011){Goossens}, {Erd{\'e}lyi}, \&
  {Ruderman}}]{goossens2011}
{Goossens}, M., {Erd{\'e}lyi}, R., \& {Ruderman}, M.~S. 2011, \ssr, 158, 289

\bibitem[{{Guo} {et~al.}(2019){Guo}, {Van Doorsselaere}, {Karampelas}, {Li},
  {Antolin}, \& {De Moortel}}]{Guo19}
{Guo}, M., {Van Doorsselaere}, T., {Karampelas}, K., {et~al.} 2019, \apj, 870,
  55

\bibitem[{{Hahn} \& {Savin}(2014)}]{Hahn2014}
{Hahn}, M. \& {Savin}, D.~W. 2014, \apj, 795, 111

\bibitem[{Harris {et~al.}(2020)Harris, Millman, van~der Walt, Gommers,
  Virtanen, Cournapeau, Wieser, Taylor, Berg, Smith, Kern, Picus, Hoyer, van
  Kerkwijk, Brett, Haldane, Fernández~del Río, Wiebe, Peterson,
  Gérard-Marchant, Sheppard, Reddy, Weckesser, Abbasi, Gohlke, \&
  Oliphant}]{numpycitation}
Harris, C.~R., Millman, K.~J., van~der Walt, S.~J., {et~al.} 2020, Nature, 585,
  357–362

\bibitem[{{Heyvaerts} \& {Priest}(1983)}]{HeyvaertPriest83}
{Heyvaerts}, J. \& {Priest}, E.~R. 1983, \aap, 117, 220

\bibitem[{{Hillier} {et~al.}(2019){Hillier}, {Barker}, {Arregui}, \&
  {Latter}}]{Hillier19}
{Hillier}, A., {Barker}, A., {Arregui}, I., \& {Latter}, H. 2019, \mnras, 482,
  1143

\bibitem[{{Hillier} \& {Polito}(2018)}]{hillier2018}
{Hillier}, A. \& {Polito}, V. 2018, \apjl, 864, L10

\bibitem[{{Hillier} {et~al.}(2020){Hillier}, {Van Doorsselaere}, \&
  {Karampelas}}]{Hillier2020}
{Hillier}, A., {Van Doorsselaere}, T., \& {Karampelas}, K. 2020, \apjl, 897,
  L13

\bibitem[{{Hollweg}(1971)}]{Hollweg71}
{Hollweg}, J.~V. 1971, \jgr, 76, 5155

\bibitem[{{Hollweg}(1978)}]{Hollweg78}
{Hollweg}, J.~V. 1978, \solphys, 56, 305

\bibitem[{{Hollweg}(1984)}]{hollweg1984}
{Hollweg}, J.~V. 1984, \apj, 277, 392

\bibitem[{{Howson} {et~al.}(2017){Howson}, {De Moortel}, \&
  {Antolin}}]{Howson17}
{Howson}, T.~A., {De Moortel}, I., \& {Antolin}, P. 2017, \aap, 607, A77

\bibitem[{{Howson} {et~al.}(2020){Howson}, {De Moortel}, \& {Reid}}]{Howson20}
{Howson}, T.~A., {De Moortel}, I., \& {Reid}, J. 2020, \aap, 636, A40

\bibitem[{Hunter(2007)}]{Huntermatplotlib}
Hunter, J.~D. 2007, Computing in Science \& Engineering, 9, 90

\bibitem[{{Iroshnikov}(1964)}]{Iroshnikov1964}
{Iroshnikov}, P.~S. 1964, \sovast, 7, 566

\bibitem[{{Jafarzadeh} {et~al.}(2017){Jafarzadeh}, {Solanki}, {Gafeira}, {van
  Noort}, {Barthol}, {Blanco Rodr{\'\i}guez}, {del Toro Iniesta}, {Gandorfer},
  {Gizon}, {Hirzberger}, {Kn{\"o}lker}, {Orozco Su{\'a}rez}, {Riethm{\"u}ller},
  \& {Schmidt}}]{Jafar17}
{Jafarzadeh}, S., {Solanki}, S.~K., {Gafeira}, R., {et~al.} 2017, \apjs, 229, 9

\bibitem[{{Jess} {et~al.}(2009){Jess}, {Mathioudakis}, {Erd{\'e}lyi},
  {Crockett}, {Keenan}, \& {Christian}}]{Jess09}
{Jess}, D.~B., {Mathioudakis}, M., {Erd{\'e}lyi}, R., {et~al.} 2009, Science,
  323, 1582

\bibitem[{{Jess} {et~al.}(2015){Jess}, {Morton}, {Verth}, {Fedun}, {Grant}, \&
  {Giagkiozis}}]{Jess15}
{Jess}, D.~B., {Morton}, R.~J., {Verth}, G., {et~al.} 2015, \ssr, 190, 103

\bibitem[{{Karampelas} \& {Van Doorsselaere}(2018)}]{Karampelas2018}
{Karampelas}, K. \& {Van Doorsselaere}, T. 2018, \aap, 610, L9

\bibitem[{{Karampelas} {et~al.}(2017){Karampelas}, {Van Doorsselaere}, \&
  {Antolin}}]{Karampelas17}
{Karampelas}, K., {Van Doorsselaere}, T., \& {Antolin}, P. 2017, \aap, 604,
  A130

\bibitem[{{Karampelas} {et~al.}(2019){Karampelas}, {Van Doorsselaere}, \&
  {Guo}}]{Karampelas19}
{Karampelas}, K., {Van Doorsselaere}, T., \& {Guo}, M. 2019, \aap, 623, A53

\bibitem[{{Kelly}(1965)}]{Kelly1965}
{Kelly}, R.~E. 1965, Journal of Fluid Mechanics, 22, 547

\bibitem[{{Kohutova} {et~al.}(2020){Kohutova}, {Verwichte}, \&
  {Froment}}]{Kohutova2020}
{Kohutova}, P., {Verwichte}, E., \& {Froment}, C. 2020, \aap, 633, L6

\bibitem[{{Kolmogorov}(1941)}]{Kolmogorov41}
{Kolmogorov}, A. 1941, Akademiia Nauk SSSR Doklady, 30, 301

\bibitem[{{Kraichnan}(1965)}]{Kraichnan1965}
{Kraichnan}, R.~H. 1965, Physics of Fluids, 8, 1385

\bibitem[{{Liu} {et~al.}(2014){Liu}, {McIntosh}, {De Moortel}, {Threlfall}, \&
  {Bethge}}]{Liu2014}
{Liu}, J., {McIntosh}, S.~W., {De Moortel}, I., {Threlfall}, J., \& {Bethge},
  C. 2014, \apj, 797, 7

\bibitem[{{Lohner}(1987)}]{Lohner87}
{Lohner}, R. 1987, Computer Methods in Applied Mechanics and Engineering, 61,
  323

\bibitem[{{Magyar} \& {Van Doorsselaere}(2016)}]{Magyar16}
{Magyar}, N. \& {Van Doorsselaere}, T. 2016, \aap, 595, A81

\bibitem[{{Magyar} {et~al.}(2019){Magyar}, {Van Doorsselaere}, \&
  {Goossens}}]{Magyar2019}
{Magyar}, N., {Van Doorsselaere}, T., \& {Goossens}, M. 2019, \apj, 882, 50

\bibitem[{{Mann} {et~al.}(1995){Mann}, {Wright}, \& {Cally}}]{Mann95}
{Mann}, I.~R., {Wright}, A.~N., \& {Cally}, P.~S. 1995, \jgr, 100, 19441

\bibitem[{{Mart{\'\i}nez-G{\'o}mez} {et~al.}(2018){Mart{\'\i}nez-G{\'o}mez},
  {Soler}, \& {Terradas}}]{David18}
{Mart{\'\i}nez-G{\'o}mez}, D., {Soler}, R., \& {Terradas}, J. 2018, \apj, 856,
  16

\bibitem[{{Mathioudakis} {et~al.}(2013){Mathioudakis}, {Jess}, \&
  {Erd{\'e}lyi}}]{Mathio13}
{Mathioudakis}, M., {Jess}, D.~B., \& {Erd{\'e}lyi}, R. 2013, \ssr, 175, 1

\bibitem[{{Matsumoto} \& {Suzuki}(2012)}]{Matsumoto12}
{Matsumoto}, T. \& {Suzuki}, T.~K. 2012, \apj, 749, 8

\bibitem[{{Mignone} {et~al.}(2007){Mignone}, {Bodo}, {Massaglia}, {Matsakos},
  {Tesileanu}, {Zanni}, \& {Ferrari}}]{Mignone07}
{Mignone}, A., {Bodo}, G., {Massaglia}, S., {et~al.} 2007, \apjs, 170, 228

\bibitem[{{Mignone} {et~al.}(2012){Mignone}, {Zanni}, {Tzeferacos}, {van
  Straalen}, {Colella}, \& {Bodo}}]{Mignone12}
{Mignone}, A., {Zanni}, C., {Tzeferacos}, P., {et~al.} 2012, \apjs, 198, 7

\bibitem[{{Morton} {et~al.}(2015){Morton}, {Tomczyk}, \& {Pinto}}]{Morton15}
{Morton}, R.~J., {Tomczyk}, S., \& {Pinto}, R. 2015, Nature Communications, 6,
  7813

\bibitem[{{Mumford} {et~al.}(2015){Mumford}, {Fedun}, \&
  {Erd{\'e}lyi}}]{Mumford2015}
{Mumford}, S.~J., {Fedun}, V., \& {Erd{\'e}lyi}, R. 2015, \apj, 799, 6

\bibitem[{{Nazarenko}(2011)}]{Nazarenko2011}
{Nazarenko}, S. 2011, {Wave Turbulence}, Vol. 825

\bibitem[{{Ng} \& {Bhattacharjee}(1997)}]{Ng1997}
{Ng}, C.~S. \& {Bhattacharjee}, A. 1997, Physics of Plasmas, 4, 605

\bibitem[{{Nocera} {et~al.}(1984){Nocera}, {Priest}, \& {Leroy}}]{Nocera84}
{Nocera}, L., {Priest}, E.~R., \& {Leroy}, B. 1984, \aap, 133, 387

\bibitem[{{Ofman} \& {Thompson}(2011)}]{Ofman11}
{Ofman}, L. \& {Thompson}, B.~J. 2011, \apjl, 734, L11

\bibitem[{{Pagano} {et~al.}(2018){Pagano}, {Pascoe}, \& {De
  Moortel}}]{Pagano18}
{Pagano}, P., {Pascoe}, D.~J., \& {De Moortel}, I. 2018, \aap, 616, A125

\bibitem[{Pope {et~al.}(2000)Pope, Eccles, Pope, \& Press}]{Pope2000}
Pope, S., Eccles, P., Pope, S., \& Press, C.~U. 2000, Turbulent Flows
  (Cambridge University Press)

\bibitem[{Powell(1994)}]{Powell94}
Powell, K.~G. 1994, An approximate Riemann solver for magnetohydrodynamics
  (that works more than one dimension), Tech. rep.

\bibitem[{Priest(2012)}]{Priest12}
Priest, E.~R. 2012, Solar magnetohydrodynamics, Vol.~21 (Springer Science \&
  Business Media)

\bibitem[{{Prokopyszyn} {et~al.}(2019){Prokopyszyn}, {Hood}, \& {De
  Moortel}}]{Prokopyszyn2019}
{Prokopyszyn}, A.~P.~K., {Hood}, A.~W., \& {De Moortel}, I. 2019, \aap, 624,
  A90

\bibitem[{{Rankin} {et~al.}(1994){Rankin}, {Frycz}, {Tikhonchuk}, \&
  {Samson}}]{Rankin94}
{Rankin}, R., {Frycz}, P., {Tikhonchuk}, V.~T., \& {Samson}, J.~C. 1994, \jgr,
  99, 21291

\bibitem[{{Roberts}(1973)}]{Roberts1973}
{Roberts}, B. 1973, Journal of Fluid Mechanics, 59, 65

\bibitem[{{Roe}(1981)}]{Roe81}
{Roe}, P.~L. 1981, Journal of Computational Physics, 43, 357

\bibitem[{{Ryutova} {et~al.}(2010){Ryutova}, {Berger}, {Frank}, {Tarbell}, \&
  {Title}}]{Ryutova10}
{Ryutova}, M., {Berger}, T., {Frank}, Z., {Tarbell}, T., \& {Title}, A. 2010,
  \solphys, 267, 75

\bibitem[{{Schekochihin} {et~al.}(2012){Schekochihin}, {Nazarenko}, \&
  {Yousef}}]{Schekochihin2012}
{Schekochihin}, A.~A., {Nazarenko}, S.~V., \& {Yousef}, T.~A. 2012, \pre, 85,
  036406

\bibitem[{Schnack(2009)}]{Schnack09}
Schnack, D. 2009, Lectures in Magnetohydrodynamics: With an Appendix on
  Extended MHD, Lecture Notes in Physics (Springer Berlin Heidelberg)

\bibitem[{{Shelyag} {et~al.}(2011){Shelyag}, {Fedun}, {Keenan}, {Erd{\'e}lyi},
  \& {Mathioudakis}}]{Shelyag2011}
{Shelyag}, S., {Fedun}, V., {Keenan}, F.~P., {Erd{\'e}lyi}, R., \&
  {Mathioudakis}, M. 2011, Annales Geophysicae, 29, 883

\bibitem[{{Shelyag} {et~al.}(2012){Shelyag}, {Mathioudakis}, \&
  {Keenan}}]{Shelyag2012}
{Shelyag}, S., {Mathioudakis}, M., \& {Keenan}, F.~P. 2012, \apjl, 753, L22

\bibitem[{{Shestov} {et~al.}(2017){Shestov}, {Nakariakov}, {Ulyanov}, {Reva},
  \& {Kuzin}}]{Shestov17}
{Shestov}, S.~V., {Nakariakov}, V.~M., {Ulyanov}, A.~S., {Reva}, A.~A., \&
  {Kuzin}, S.~V. 2017, \apj, 840, 64

\bibitem[{{Shoda} {et~al.}(2018){Shoda}, {Yokoyama}, \& {Suzuki}}]{Shoda2018}
{Shoda}, M., {Yokoyama}, T., \& {Suzuki}, T.~K. 2018, \apj, 853, 190

\bibitem[{{Smith} {et~al.}(2007){Smith}, {Tsiklauri}, \& {Ruderman}}]{Smith07}
{Smith}, P.~D., {Tsiklauri}, D., \& {Ruderman}, M.~S. 2007, \aap, 475, 1111

\bibitem[{{Soler} \& {Terradas}(2015)}]{Soler15}
{Soler}, R. \& {Terradas}, J. 2015, \apj, 803, 43

\bibitem[{{Soler} {et~al.}(2019){Soler}, {Terradas}, {Oliver}, \&
  {Ballester}}]{Soler19}
{Soler}, R., {Terradas}, J., {Oliver}, R., \& {Ballester}, J.~L. 2019, \apj,
  871, 3

\bibitem[{{Soler} {et~al.}(2021){Soler}, {Terradas}, {Oliver}, \&
  {Ballester}}]{Soler20}
{Soler}, R., {Terradas}, J., {Oliver}, R., \& {Ballester}, J.~L. 2021, \apj, in
  press

\bibitem[{{Soler} {et~al.}(2010){Soler}, {Terradas}, {Oliver}, {Ballester}, \&
  {Goossens}}]{soler2010}
{Soler}, R., {Terradas}, J., {Oliver}, R., {Ballester}, J.~L., \& {Goossens},
  M. 2010, \apj, 712, 875

\bibitem[{{Srivastava} \& {Dwivedi}(2017)}]{Srivastava17}
{Srivastava}, A.~K. \& {Dwivedi}, B.~N. 2017, Journal of Astrophysics and
  Astronomy, 38, 61

\bibitem[{{Srivastava} {et~al.}(2017){Srivastava}, {Shetye}, {Murawski},
  {Doyle}, {Stangalini}, {Scullion}, {Ray}, {W{\'o}jcik}, \&
  {Dwivedi}}]{Srivastava2017}
{Srivastava}, A.~K., {Shetye}, J., {Murawski}, K., {et~al.} 2017, Scientific
  Reports, 7, 43147

\bibitem[{Stix(2012)}]{stix12}
Stix, M. 2012, The sun: an introduction (Springer Science \& Business Media)

\bibitem[{{Terradas} {et~al.}(2008){Terradas}, {Andries}, {Goossens},
  {Arregui}, {Oliver}, \& {Ballester}}]{Terradas08}
{Terradas}, J., {Andries}, J., {Goossens}, M., {et~al.} 2008, \apjl, 687, L115

\bibitem[{{Terradas} \& {Arregui}(2018)}]{terradasarregui2018}
{Terradas}, J. \& {Arregui}, I. 2018, Research Notes of the American
  Astronomical Society, 2, 196

\bibitem[{{Terradas} {et~al.}(2018){Terradas}, {Magyar}, \& {Van
  Doorsselaere}}]{Terradas18}
{Terradas}, J., {Magyar}, N., \& {Van Doorsselaere}, T. 2018, \apj, 853, 35

\bibitem[{{Terradas} \& {Ofman}(2004)}]{Terradas04}
{Terradas}, J. \& {Ofman}, L. 2004, \apj, 610, 523

\bibitem[{{Terradas} {et~al.}(2006){Terradas}, {Oliver}, \&
  {Ballester}}]{terradas06}
{Terradas}, J., {Oliver}, R., \& {Ballester}, J.~L. 2006, \apj, 642, 533

\bibitem[{Tikhonchuk {et~al.}(1995)Tikhonchuk, Rankin, Frycz, \&
  Samson}]{Tikhonchuk95}
Tikhonchuk, V.~T., Rankin, R., Frycz, P., \& Samson, J.~C. 1995, Physics of
  Plasmas, 2, 501

\bibitem[{{van Ballegooijen} {et~al.}(2011){van Ballegooijen}, {Asgari-Targhi},
  {Cranmer}, \& {DeLuca}}]{vanBallegooijen2011}
{van Ballegooijen}, A.~A., {Asgari-Targhi}, M., {Cranmer}, S.~R., \& {DeLuca},
  E.~E. 2011, \apj, 736, 3

\bibitem[{{van Ballegooijen} {et~al.}(2017){van Ballegooijen}, {Asgari-Targhi},
  \& {Voss}}]{vanBallegooijen2017}
{van Ballegooijen}, A.~A., {Asgari-Targhi}, M., \& {Voss}, A. 2017, \apj, 849,
  46

\bibitem[{{Van Damme} {et~al.}(2020){Van Damme}, {De Moortel}, {Pagano}, \&
  {Johnston}}]{Vandamme20}
{Van Damme}, H.~J., {De Moortel}, I., {Pagano}, P., \& {Johnston}, C.~D. 2020,
  \aap, 635, A174

\bibitem[{Virtanen {et~al.}(2020)Virtanen, Gommers, Oliphant, Haberland, Reddy,
  Cournapeau, Burovski, Peterson, Weckesser, Bright, {van der Walt}, Brett,
  Wilson, Millman, Mayorov, Nelson, Jones, Kern, Larson, Carey, Polat, Feng,
  Moore, {VanderPlas}, Laxalde, Perktold, Cimrman, Henriksen, Quintero, Harris,
  Archibald, Ribeiro, Pedregosa, {van Mulbregt}, \& {SciPy 1.0
  Contributors}}]{scipycitation}
Virtanen, P., Gommers, R., Oliphant, T.~E., {et~al.} 2020, Nature Methods, 17,
  261

\bibitem[{{Wedemeyer-B{\"o}hm} {et~al.}(2012){Wedemeyer-B{\"o}hm}, {Scullion},
  {Steiner}, {Rouppe van der Voort}, {de La Cruz Rodriguez}, {Fedun}, \&
  {Erd{\'e}lyi}}]{Wedemeyer2012}
{Wedemeyer-B{\"o}hm}, S., {Scullion}, E., {Steiner}, O., {et~al.} 2012, \nat,
  486, 505

\bibitem[{{Zaqarashvili}(2003)}]{Zaqarashvili2003}
{Zaqarashvili}, T.~V. 2003, \aap, 399, L15

\bibitem[{{Zaqarashvili} {et~al.}(2015){Zaqarashvili}, {Zhelyazkov}, \&
  {Ofman}}]{Zaqarashvili2015}
{Zaqarashvili}, T.~V., {Zhelyazkov}, I., \& {Ofman}, L. 2015, \apj, 813, 123

\end{thebibliography}
\end{document}